\documentclass[showpacs,aps,prd,nofootinbib,showkeys,unsortedaddress,twocolumn,floatfix]{revtex4-1}
\pdfoutput=1

\usepackage{makecell}
\usepackage{bm}
\usepackage{amsmath}
\usepackage{graphicx}
\usepackage[usenames,dvipsnames]{color}
\definecolor{darkblue}{RGB}{0,0,196}
\usepackage[colorlinks=true,linkcolor=darkblue,citecolor=darkblue,urlcolor=darkblue]{hyperref}

\newcommand{\beq}{\begin{equation}}
\newcommand{\eeq}{\end{equation}}
\newcommand{\bqa}{\begin{eqnarray}}
\newcommand{\eqa}{\end{eqnarray}}

\begin{document}

\title{Dilepton production from the quark-gluon plasma using (3+1)-dimensional anisotropic dissipative hydrodynamics}

\author{Radoslaw Ryblewski}
\affiliation{The H. Niewodnicza\'nski Institute of Nuclear Physics, Polish Academy of Sciences, PL-31342 Krak\'ow, Poland} 

\author{Michael Strickland} 
\affiliation{Department of Physics, Kent State University, Kent, OH 44242 United States}

\begin{abstract}
We compute dilepton production from the deconfined phase of the quark-gluon plasma using leading-order (3+1)-dimensional anisotropic hydrodynamics.  The anisotropic hydrodynamics equations employed describe the full spatiotemporal evolution of the transverse temperature, spheroidal momentum-space anisotropy parameter, and the associated three-dimensional collective flow of the matter.  The momentum-space anisotropy is also taken into account in the computation of the dilepton production rate, allowing for a self-consistent description of dilepton production from the quark-gluon plasma.  For our final results, we present predictions for high-energy dilepton yields as a function of invariant mass, transverse momentum, and pair rapidity.  We demonstrate that high-energy dilepton production is extremely sensitive to the assumed level of initial momentum-space anisotropy of the quark-gluon plasma.  As a result, it may be possible to experimentally constrain the early-time momentum-space anisotropy of the quark-gluon plasma generated in relativistic heavy ion collisions using high-energy dilepton yields.
\end{abstract}


\maketitle 

\section {Introduction}
\label{sec:intro}

The degree to which the quark and gluon distributions of the partons comprising the quark-gluon plasma (QGP) generated in relativistic heavy-ion collisions are momentum-space isotropic in the local rest frame (LRF) is currently an open question.  There have been a number of theoretical studies that have attempted to address this question using both perturbative QCD and the AdS/CFT framework (see Ref.~\cite{Strickland:2013uga} for a recent review).  Ideally, however, one would like to have an experimental observable that could provide constraints on the degree of isotropy during the early stages of the QGP's lifetime and perhaps, in addition, the subsequent approach towards isotropy.  

In principle, electromagnetic emissions are the ideal observable for studying the early-time dynamics of the QGP since they are weakly coupled to the plasma ($\alpha \ll \alpha_s$).  In addition, due to the fact that the QGP is initially hot and then cools, high-energy ($E \gtrsim 2$ GeV) production is dominated by early times when the system is in the QGP phase, while low-energy ($E \lesssim 2$ GeV) production receives significant contributions from late-time emissions when the system returns to the hadronic phase.  This simple picture is complicated by the fact that there is a temperature distribution in the QGP, with the edges of the system being best described using hadronic degrees of freedom, however, since these regions are rather dilute and small in relative volume, the total radiation from this region is small compared to that produced from the central region.  The two primary electromagnetic observables studied in heavy-ion collisions are real photons and dileptons produced via decay of virtual photons.

In this paper we focus on dilepton production from the deconfined phase of the QGP's lifetime.  The study of dilepton production from the QGP has a long history, see e.g. Refs.~\cite{Shuryak:1978ij,Domokos:1980ba,Kajantie:1981wg,Kajantie:1986dh,Kapusta:1992uy,Strickland:1994rf,
Rapp:1999ej,Rapp:1999zw,Rapp:2000pe,Arnold:2001ba,Arnold:2001ms,Aurenche:2002wq,Arnold:2002ja,Dusling:2008xj,Vujanovic:2013jpa,Endres:2014zua}.  For recent reviews, see also Refs.~\cite{Rapp:2013nxa,Sakaguchi:2014ewa}.  Herein we focus on the effect of LRF momentum-space anisotropies on dilepton production.  This work is an extension of previous studies performed in Refs.~\cite{Mauricio:2007vz,Martinez:2008di} to include a realistic bulk evolution using the framework of anisotropic hydrodynamics \cite{Martinez:2010sc,Florkowski:2010cf,Ryblewski:2010bs,Martinez:2010sd,Ryblewski:2011aq,Florkowski:2011jg,Martinez:2012tu,Ryblewski:2012rr,Florkowski:2012as,Florkowski:2013uqa,Ryblewski:2013jsa,Florkowski:2013lza,Bazow:2013ifa,Tinti:2013vba,Florkowski:2014bba,Florkowski:2014txa,Nopoush:2014pfa,Denicol:2014mca,Nopoush:2014qba,Tinti:2014yya}.  For a recent review of the motivation for and methods used to obtain the anisotropic hydrodynamics equations and solve them numerically, we refer the reader to Ref.~\cite{Strickland:2014pga}.  

Herein, we make use of the anisotropic hydrodynamics equations obtained from the zeroth and first moments of the Boltzmann equation with the collisional kernel treated in the relaxation-time approximation and describe the (3+1)-dimensional evolution of the QGP using these equations.  The resulting dynamical equations describe the full spatiotemporal evolution of the transverse temperature $\Lambda$ and spheroidal momentum-space anisotropy parameter $\xi$ \cite{Romatschke:2003ms,Romatschke:2004jh}.  The (3+1)-dimensional framework allows both $\Lambda$, $\xi$, and the associated flow velocities to depend arbitrarily on the transverse coordinates, spatial rapidity, and longitudinal proper-time, however, herein we restrict ourselves to smooth Glauber-like initial conditions. 

The study presented herein is similar in spirit to prior studies of dilepton production using viscous hydrodynamics \cite{Dusling:2008xj,Vujanovic:2013jpa}.  In these works, however, the authors employed the standard viscous hydrodynamic linearization around an an isotropic thermal background.   Our work goes beyond these studies by linearizing around anisotropic background and, as a result, we are able to better describe  early-time dilepton production and dilepton production near the transverse and longitudinal edges of the QGP.  In addition, high-momentum dilepton production is treated in a more reliable manner since the anisotropic one-particle distribution function used to compute the dilepton rates is positive definite at all points in momentum space.  We demonstrate that high-energy dilepton production is extremely sensitive to the assumed level of initial momentum-space anisotropy of the quark-gluon plasma. As a result, it may be possible to experimentally constrain the early-time momentum-space anisotropy of the quark-gluon plasma generated in relativistic heavy-ion collisions using high-energy dilepton yields.

The structure of our paper is as follows.   In Sec.~\ref{sec:dileptonrate}, we review the calculation of the leading-order dilepton production rate in an anisotropic QGP.  In Sec.~\ref{sec:spectra} we describe how one calculates the dilepton spectra including the effect of transverse and longitudinal expansion.  In Sec.~\ref{sec:hydro} we present the setup for the anisotropic hydrodynamics evolution, the resulting (3+1)-dimensional dynamical equations, the equation of state employed, and the initial conditions used.  In Sec.~\ref{sec:results} we present our final numerical results for the dilepton yields as a function of invariant mass, transverse momentum, and pair rapidity using fixed initial conditions and fixed final multiplicity.  We present our conclusions and an outlook for the future in Sec.~\ref{sec:conc}.  We collect information about particle production from (3+1)-dimensional anisotropic hydrodynamics and compare this with Israel-Stewart viscous hydrodynamics in App.~\ref{sec:app1}.

\section{Dilepton Rate in anisotropic plasma}
\label{sec:dileptonrate}

We begin by reviewing the derivation of the dilepton emission rate for an anisotropic plasma starting from relativistic kinetic theory. We follow the methodology presented originally in Ref.~\cite{Martinez:2008di}. The resulting formulas will be subsequently used in Sec.~\ref{sec:spectra} to calculate the differential dilepton spectra using anisotropic hydrodynamics framework.

The dilepton emission rate is defined as the number of dilepton pairs produced per eight-dimensional phase-space volume
\begin{equation}
\frac{d R^{l^+l^-}}{d^4\!P}\equiv\frac{dN^{l^+l^-}}{d^4\!X d^4\!P} \, ,
\label{rate}
\end{equation}
where $X^{\mu} = (t,{\bf x})$ and $P^{\mu} = (E,{\bf p})$ are the four-position and four-momentum, respectively.
Based on relativistic kinetic theory,\footnote{The same result can be obtained using standard finite temperature field theory techniques.} at leading order in the electromagnetic coupling, ${\cal O}(\alpha^2)$, the dilepton emission rate follows from
\begin{eqnarray}
\frac{d R^{l^+l^-}}{d^4\!P} &=& \int \frac{d^3{\bf p}_1}{(2\pi)^3}\,\frac{d^3{\bf p}_2}{(2\pi)^3}\, f_q({\bf p}_1)\,f_{\bar{q}}({\bf p}_2)\, \nonumber \\ 
&& \times\, v_{q\bar{q}}\,\sigma^{l^+l^-}_{q\bar{q}}\, \delta^{(4)}(P^\mu-p_1^\mu-p_2^\mu) \, ,
\label{kineticrate}
\end{eqnarray}
where $f_{q ({\bar q})}$ is the phase-space distribution function of quarks (anti-quarks),\footnote{From now on we assume that $f_{\bar q}=f_q$.} $\it{v}_{q\bar{q}}$ is the relative velocity between the quark and the anti-quark
\begin{equation}
v_{q\bar{q}} \equiv \frac{\sqrt{{\bf p}_1 \cdot {\bf p}_2 - m_q^2}}{2E_{{\bf p}_1} 2E_{{\bf p}_2}} \, ,
\label{relvelocity}
\end{equation}
and $\sigma^{l^+l^-}_{q\bar{q}}$ is the total cross section for the leading-order quark--anti-quark annihilation process, $q + \bar{q} \to \gamma^\ast \to l^+ + l^-$
\begin{equation}
\sigma^{l^+l^-}_{q\bar{q}} = \frac{4\pi}{3} \frac{\alpha^2}{M^2} 
		\left(1 + \frac{2 m_l^2}{M^2}\right) 
		\left(1 - \frac{4 m_l^2}{M^2}\right)^{1/2} .
\label{crosssection}
\end{equation}
Henceforth, we will consider only high-energy dilepton pairs with 
invariant energies much greater than the lepton masses, $M \gg m_l$. Therefore, we will ignore lepton mass corrections appearing in Eq.~(\ref{crosssection}) and simply take $m_l = 0$.

Ultra-relativistic heavy-ion collisions are special in the sense that the matter created in such events is undergoing rapid expansion along the longitudinal (beam) direction. At the same time, the transverse expansion is initially relatively quite slow. One can show that this phenomenon inevitably leads to the presence of large momentum-space anisotropies in the phase-space distribution of the matter. The simplest form for the distribution function that can be used to describe this situation is a generalization of an isotropic phase-space distribution which is squeezed or stretched along one direction in momentum space, defined by $\hat{\bf n}$, with a parameter $-1 < \xi < \infty$, which describes the type and strength of the momentum-space anisotropy.  In this case, the one-particle distribution function for the quarks and anti-quarks may be described at leading order by the following spheroidal ``Romatschke-Strickland'' form \cite{Romatschke:2003ms,Romatschke:2004jh}
\begin{equation}
f_{q ({\bar q})}({\bf p},\xi,\Lambda)\equiv f^{\rm iso}_{q ({\bar 
q})}(\sqrt{{\bf p}^2+\xi({\bf p\cdot \hat{n}})^2},\Lambda) \, ,
\label{distansatz}
\end{equation}
where $\Lambda$ is a transverse-momentum scale and $\xi$ is the anisotropy parameter introduced above. In the limiting case where $\xi=0$, Eq.~(\ref{distansatz}) reduces to the standard isotropic distribution function. When $\xi=0$, $\Lambda$ can be identified with the equilibrium temperature $T$ of the system.  Herein, we will take $f^{\rm iso}_{q ({\bar q})}$ to be a Fermi-Dirac distribution function $f^{\rm iso}_{q ({\bar q})}(E,T) = [\,\exp(E/T)+1\,]^{-1}$. 

\begin{figure*}[t]
\centerline{
\includegraphics[width=0.315\linewidth]{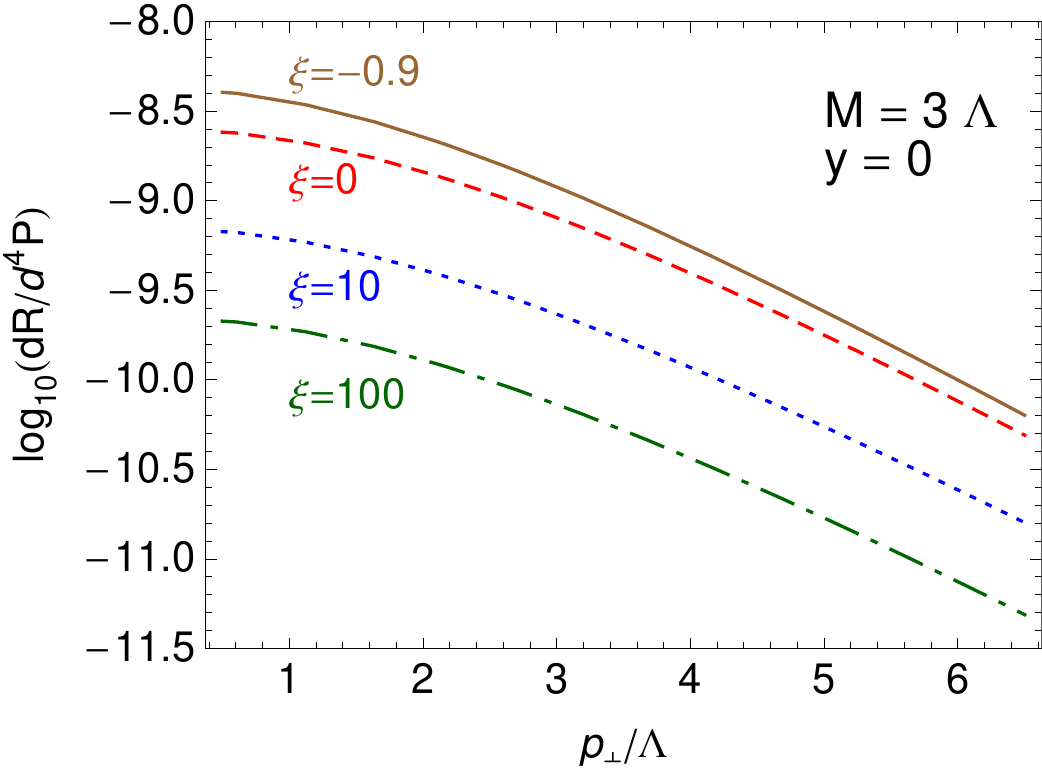}\hspace{3mm}
\includegraphics[width=0.315\linewidth]{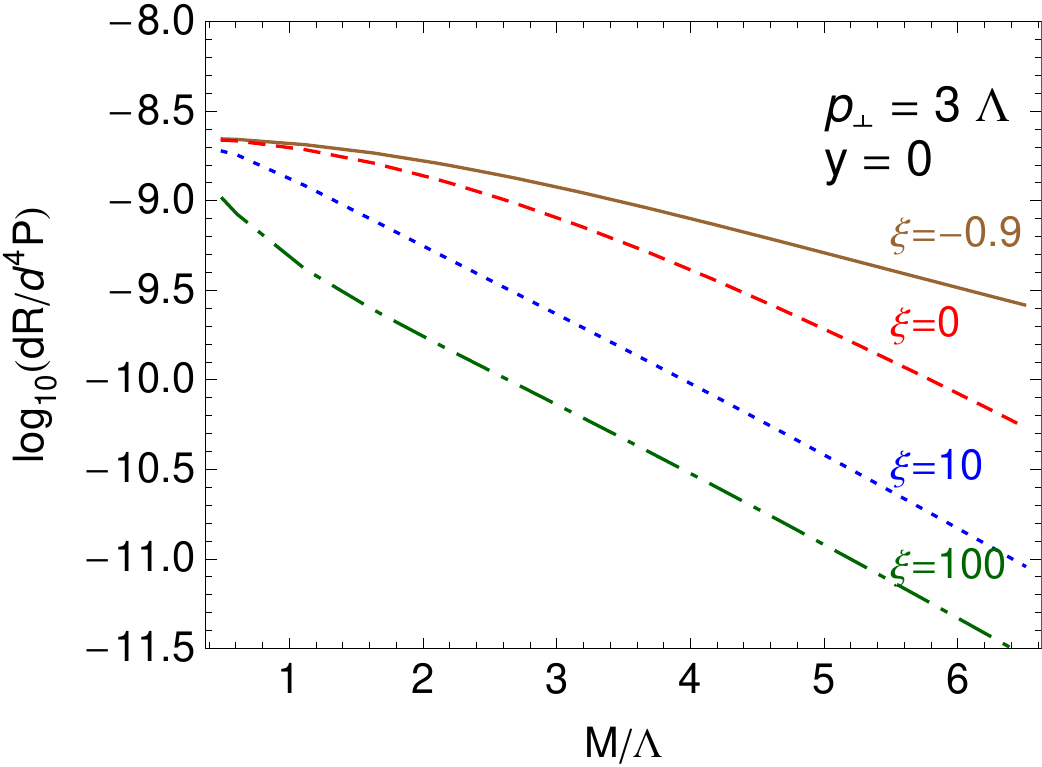}\hspace{3mm}
\includegraphics[width=0.315\linewidth]{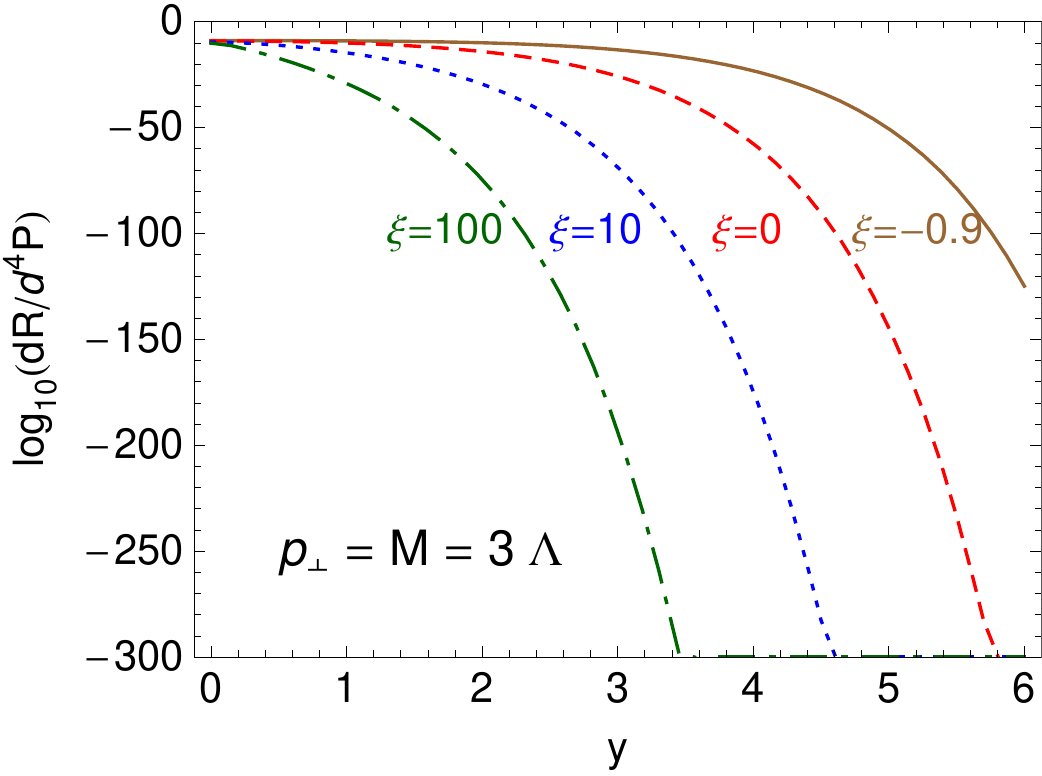}
}
\caption{
The dilepton emission rate as a function of transverse momentum (left), invariant mass (center) and rapidity (right). For the transverse momentum dependence (left) we fixed $M / \Lambda$= 3 and $y$=0, for the invariant mass dependence (center) we fixed $p_\perp / \Lambda$= 3 and $y$=0, for rapidity dependence we fixed $p_\perp / \Lambda = M / \Lambda$= 3.
}
\label{diffrate}
\end{figure*}

Using the Dirac delta function in Eq.~(\ref{kineticrate}), one can immediately perform the ${\bf p}_2$ integration to obtain

\begin{eqnarray}
\frac{d R^{l^+l^-}}{d^4\!P} &=& \frac{5\alpha^2}{72\pi^5}\int\frac{d^3{\bf p}_1}{ E_{{\bf p}_1} E_{{\bf p}_2}}\, f_q({\bf p}_1,\Lambda, {\bf \xi})\, f_{\bar{q}}({\bf p}_2,\Lambda, {\bf \xi}) \nonumber \\
&&\times\, \delta (E-\! E_{{\bf p}_1}\! \! -\! E_{{\bf p}_2}) \Biggr|_{{\bf p}_2={\bf P}-{\bf p}_1} \, .
\label{ratemiddle}
\end{eqnarray}
To proceed, we parameterize the remaining three-momenta using spherical coordinates with the $z$-axis defined by the direction of anisotropy ${\bf \hat{n}}$,
\begin{eqnarray}
{\bf p}_1 &=&
p_1(\sin\theta_{p_1}\cos\phi_{p_1},\sin\theta_{p_1}\sin\phi_{p_1},\cos\theta_{p_1}),\nonumber
\\
{\bf P} &=&
P(\sin\theta_{P}\cos\phi_{P},\sin\theta_{P}\sin\phi_{P},\cos\theta_{P}).
\end{eqnarray}
In this way we may rewrite the remaining delta function in (\ref{ratemiddle}) in the form
\begin{equation}
\delta (E-\! E_{{\bf p}_1}\! \! -\! E_{{\bf p}_2}) =2\,(E-p_1)\frac{\Theta(\chi)}{\sqrt{\chi}}\sum_i^2\delta (\phi_i-\phi_{p_1}) \, ,
\label{delta}
\end{equation}
where
\begin{eqnarray}
\chi &\equiv& (2 p_1 P \sin\theta_P\sin\theta_{p_1})^2 \nonumber \\
&& \hspace{1cm} -[2p_1(E-P\cos\theta_P\cos\theta_{p_1})-M^2]^2 \, .
\end{eqnarray}
The angles $\phi_i$ are calculated as the two possible solutions to the equation
\begin{equation}
\cos\,(\phi_i-\phi_{p_1})=\frac{2 p_1 (E-P\cos\theta_P\cos\theta_{p_1})-M^2}{2 p_1 P \sin\theta_P\sin\theta_{p_1}} \, .
\label{coseno}
\end{equation}
After these substitutions, we arrive at our final result for the dilepton emission rate 
\begin{widetext}
\begin{eqnarray}
\frac{d R^{l^+l^-}}{d^4\!P}\!\! &=&\!
\frac{5\alpha^2}{18\pi^5}\int_{-1}^1 \!d(\cos\theta_{p_1})
\!\int_{a_+}^{a_-}\!\! \frac{p_1 dp_1}{\sqrt{\chi}}\, f_q\!\left({p_1\sqrt{\!1+\!\xi\cos^2\theta_{p_1}}},\Lambda\right)
\nonumber \\
&& \hspace{4cm} \times f_{\bar{q}}\!\left(\sqrt{{(E\!-\!p_1)^2+\xi(p_1\cos\theta_{p_1}\!\!-P\cos\theta_P)^2}},\Lambda\right),
\label{kineticratefinal}
\end{eqnarray}
\end{widetext}
with
\begin{eqnarray}
a_{\pm}&\equiv&\frac{M^2}{2(E-P\cos (\theta_P\pm\theta_{p_1}))} \, .
\end{eqnarray}
In order to evaluate the dilepton emission rate (\ref{kineticratefinal}) it is necessary to perform the remaining two integrations numerically. In Fig.~\ref{diffrate} we show the resulting dilepton emission rate as a function of transverse momentum (left) and invariant mass (center), both scaled by $\Lambda$, and rapidity (right) for various values of anisotropy parameter $\xi \in \{-0.9, 0, 10, 100\}$ denoted by brown solid, red dashed, blue dotted and green dot-dashed lines, respectively. One can see that the production rate decreases (increases) due to increasing (decreasing) $\xi$.   We note, however, that this is primarily due to the fact that increasing $\xi$ for fixed $\Lambda$ results in a lower plasma density.  In order, to properly assess the impact of anisotropies on the production, one has to fold these rates together with a realistic model of the full spatiotemporal evolution of both $\xi$ and $\Lambda$.

\section {Dilepton spectra}
\label{sec:spectra}

Our final goal is to study the impact of space-time dependent anisotropies in the system on the dilepton differential spectra. In this way we hope to probe the early stages of the quark-gluon plasma, where the anisotropies are expected to be the largest. In order to do this one must include in Eq.~(\ref{kineticratefinal}) the space-time dependence of $\Lambda$ and $\xi$ using some hydrodynamic model and then integrate over the entire space-time volume (which contains the quark-gluon plasma phase) and the appropriate momenta/invariant mass cuts for the dilepton pairs. For this purpose, we parametrize the pair four-momentum in the standard way,
\begin{equation}
p^{\mu}=(m_{\perp} \cosh y, p_{\perp} \cos \phi_p, p_{\perp} \sin \phi_p, m_{\perp} \sinh y) \, ,
\label{mompar}
\end{equation}
where $m_{\perp}\equiv\sqrt{M^2 + p_{\perp}^2}$ defines the transverse mass and $y \equiv 1/2 \ln\!\left[(E+p_\parallel)/(E-p_\parallel)\right]$ is the momentum-space rapidity. Above, we used $p_{\perp}$, $p_{\parallel}$, and $\phi_p$ to denote transverse momentum, longitudinal momentum, and momentum azimuthal angle, respectively.

One can also use the usual Milne hyperbolic parametrization of space-time which is convenient for describing heavy-ion collisions within the relativistic hydrodynamics framework
\begin{equation}
x^{\mu}=(\tau \cosh \varsigma, {\bf x}_\perp, \tau \sinh \varsigma) \, .
\label{spacepar}
\end{equation}
In Eq.~(\ref{spacepar}), we used $\tau\equiv\sqrt{t^2 - z^2}$ and $\varsigma \equiv \tanh^{-1} (z/t)$ to denote the longitudinal proper time and the space-time rapidity, respectively. With these parameterizations, the differential measures for four-momentum  and space-time are $d^4\!P=M dM \,dy \,p_{\perp} dp_\perp \,d\phi_p$ and $d^4\!X=\tau d\tau \,d\varsigma \, d^2 x_\perp$, respectively. This allows us to calculate the invariant mass and transverse momentum differential spectra using
\begin{subequations}
\begin{align}
\frac{dN^{l^+l^-}}{MdMdy}&=\int_{p_\perp^{\rm min}}^{p_\perp^{\rm max}} \!\!\!\!\! p_\perp dp_\perp
   \int_{0}^{2\pi} \!\! d\phi_p
   \int \!\! d^4\!X \frac{dR^{l^+l^-}}{d^4\!P} \, , \label{Mspectrum}\\
\frac{dN^{l^+l^-}}{p_\perp dp_\perp dy}&=\int_{M^{\rm min}}^{M^{\rm max}} \!\!\!\!\!\!\!\! M dM
   \int_{0}^{2\pi} \!\! d\phi_p 
   \int \!\! d^4\!X \frac{dR^{l^+l^-}}{d^4\!P} \, ,\label{pTspectrum}
\end{align}
\label{spectrumeqs}
\end{subequations}
respectively, where the integration ranges $p_\perp^{\rm min}$, $p_\perp^{\rm max}$ and $M^{\rm min}$, $M^{\rm max}$ will be specified later according to the appropriate physical/experimental cuts. The integration over the space-time volume is performed only in the deconfined quark-gluon plasma stage. In practice, we only include contributions from regions that have an effective temperature that is higher than a critical temperature, i.e. $T \equiv {\cal R}^{1/4}(\xi)\Lambda > T_c$ with ${\cal R}(\xi)$ defined in Eq.~(\ref{R}).  In all results shown herein, we assume $T_c = 175$ MeV. We will assume that when the system reaches $T_c$, all medium emission stops.  We do not take into account the emission from the mixed/hadronic phase at late times since the kinematic regime we study (high $M$ and $p_\perp$) is dominated by early-time high-energy dilepton emission. Due to the large uncertainty connected with the correct value of critical temperature existing in the literature we also checked that the results obtained here are almost completely independent of the choice of $T_c$ in the range $150 - 200$ MeV. 

Equations~(\ref{Mspectrum}) and (\ref{pTspectrum}) are evaluated in the center of mass of the colliding nuclei (LAB) frame while the dilepton emission rate is calculated in the local rest frame (LRF) of the emitting region. Therefore, before evaluating Eqs.~(\ref{spectrumeqs}) we have to boost the LAB frame momentum $p^{\mu}$ to the LRF of the fluid cell using $p^{\prime \mu} = \Lambda^{\mu\,\,}_{\,\,\nu}\,\, p^{\nu}$, where the Lorentz boost tensor
\begin{widetext}
\begin{equation}
\Lambda^{\mu\,\,}_{\,\,\nu}(u^{\mu}) \equiv \left(
\begin{array}{rrrr}
\gamma &  -\gamma v_x  &  -\gamma v_y  &  -\gamma v_z \\
-\gamma v_x & 1 + (\gamma - 1) \frac{v_x^2}{v^2}    &     (\gamma - 1) \frac{v_x v_y}{v^2}  &     (\gamma - 1) \frac{v_x v_z}{v^2} \\
-\gamma v_y &     (\gamma - 1) \frac{v_x v_y}{v^2}  & 1 + (\gamma - 1) \frac{v_y^2}{v^2}    &     (\gamma - 1) \frac{v_y v_z}{v^2} \\
-\gamma v_z &     (\gamma - 1) \frac{v_x v_z}{v^2}  &     (\gamma - 1) \frac{v_y v_z}{v^2}  & 1 + (\gamma - 1) \frac{v_z^2}{v^2}
\end{array} \right),
\label{boost}
\end{equation}
\end{widetext}
depends on the four-velocity of the fluid element $u^{\mu}(x^{\mu}) \equiv \gamma (1, v_x, v_y, v_z)$, where $\gamma \equiv 1 / \sqrt{1 - v^2}$ and $v \equiv \sqrt{v_z^2 + v_y^2 +v_z^2}$. One can easily check that, as expected, $u_{\rm LRF}^{\mu} = \Lambda^{\mu\,\,}_{\,\,\nu}\,\, u^{\nu} = (1, 0, 0, 0)$.  Making use of Eq.~(\ref{kineticratefinal}) in Eqs.~(\ref{spectrumeqs}), we obtain the dilepton spectra including the effect of a space-time-dependent momentum anisotropy. 

\section {Hydrodynamic evolution}
\label{sec:hydro}

As mentioned above, in order to make predictions for the differential dilepton spectra expected to be produced from the QGP phase, one must integrate over the full space-time history of the QGP.  For this purpose, we use anisotropic hydrodynamics.  Anisotropic hydrodynamics reduces to second-order viscous hydrodynamics in the limit of small anisotropy \cite{Tinti:2014yya}, but reproduces the dynamics of the QGP more reliably when there are large momentum-space anisotropies.

\subsection{(3+1)-dimensional anisotropic hydrodynamics}
\label{ssec:3+1}

In this paper, we assume that the system created during the collision of the heavy ions evolves through a non-equilibrium state and that the quark and anti-quark one-particle distribution functions are well approximated by Eq.~(\ref{distansatz}) both at early times and late times.  At the same time, we assume that, although the system is highly anisotropic it may still be, to good approximation, described using hydrodynamic-like degrees of freedom, such as energy density and pressures.\footnote{This assumption has been tested elsewhere by comparing the predictions of anisotropic hydrodynamics to exact solutions of the Boltzmann equation in a variety of special cases \cite{Florkowski:2013lza,Florkowski:2013lya,Bazow:2013ifa,Florkowski:2014sfa,Florkowski:2014sda,Denicol:2014xca,Denicol:2014tha,Nopoush:2014qba}.  These studies found that anisotropic hydrodynamics provides the most accurate description of both the early and late time behavior of QGP dynamics.}   In this way, the detailed microscopic description of the system can be replaced by an effective description which realizes simple physical laws, such as conservation of energy and momentum. In the following, we will present the framework of leading-order anisotropic hydrodynamics \cite{Florkowski:2010cf,Martinez:2010sc,Ryblewski:2010bs,Martinez:2010sd,Ryblewski:2011aq,Martinez:2012tu,Ryblewski:2012rr} which is designed to describe a potentially highly-anisotropic plasma by assuming that its distribution function is, to good approximation, expressible in the form given by Eq.~(\ref{distansatz}).

At leading order, one can derive the equations of motion of the anisotropic system starting from kinetic theory assuming that the distribution function of the system is known. This can be done by taking moments of the Boltzmann kinetic equation with the collision term treated in the relaxation-time approximation (RTA)
\begin{eqnarray}
p^\mu \partial_\mu f &=& \frac{p^\mu u_\mu}{\tau_{\rm eq}}(f_{\rm iso}-f)\,,
\label{eq:boltzmanneq}
\end{eqnarray}
where $\tau_{\rm eq}$ is the microscopic relaxation time which can depend on position and time.
Taking the first moment of the Boltzmann equation results in the energy-momentum conservation equation
\begin{eqnarray}
\partial_\mu T^{\mu \nu} &=& 0 \, . \label{enmomcon}
\end{eqnarray}
Taking the zeroth moment of the Boltzmann equation results in the particle production equation
\begin{eqnarray}
\partial_\mu N^{\mu} &=& u_\mu \frac{N^{\mu}_{\rm eq}-N^{\mu}}{\tau_{\rm eq}} \, . \label{partprod}
\end{eqnarray}
At leading order, the energy-momentum tensor has the form typical for a spheroidally anisotropic system
\begin{equation}
T^{\mu \nu} = \left( \varepsilon  + P_{\perp}\right) u^{\mu}u^{\nu} - P_{\perp} \, g^{\mu\nu} - (P_{\perp} - P_{\parallel}) z^{\mu}z^{\nu} \, ,
\label{Taniso}
\end{equation}
and the particle flux is defined in the standard manner
\begin{eqnarray}
N^{\mu}_{\rm eq} &=&  n_{\rm eq} \, u^\mu \, .
\label{Naniso}
\end{eqnarray}
In Eqs.~(\ref{Taniso}) and (\ref{Naniso}) $\varepsilon$, $n$, $P_{\parallel}$, and $P_{\perp}$ stand for energy density, particle density, longitudinal pressure, and transverse pressure, respectively. The four-vector $z^\mu$ is orthogonal to $u^{\mu}$ and in the LRF points in the longitudinal direction (identified with the direction of the anisotropy in the system, $\bf \hat n$) \cite{Martinez:2012tu}.

Equations (\ref{enmomcon}) and (\ref{partprod}) provide a set of five independent partial differential equations
\begin{eqnarray}
D_u \varepsilon &=& - \left( \varepsilon+P_\perp \right) \theta_u 
+ \left( P_\perp-P_\parallel \right) u_\nu  D_z z^\nu \, , \hspace{6mm}
\label{enmomconU} \\
D_z P_\parallel &=&  \left( P_\perp-P_\parallel \right) \theta_z  + \left( \varepsilon+ P_\perp \right) z_\nu  D_u u^\nu \, , 
\label{enmomconeqs}\\
D_u u_{\perp} &=& - \frac{u_{\perp}}{\varepsilon + P_{\perp}} \Bigg[ \frac{ {\bf u}_\perp \cdot {\bf \nabla}_\perp P_{\perp}}{u_\perp^2} \nonumber \\
& & \hspace{7mm} +  D_u P_{\perp} +  (P_{\perp} - P_{\parallel}) u_\nu D_z z^\nu \frac{}{} \Bigg] ,
\label{HydEqEuler1}\\
D_u \left( \frac{u_x}{u_y} \right) &=& \frac{1}{u_y^2 (\varepsilon + P_{\perp})} \left( u_x \partial_y - u_y \partial_x \right)P_{\perp} \, ,
\label{HydEqEuler2}
\end{eqnarray}
and 
\begin{equation}
\frac{D_u \xi}{2 (1+\xi)} - \frac{3 D_u \Lambda}{\Lambda} = \theta_u + \frac{1}{\tau_{\rm eq}} \left[1 - {\cal R}^{3/4}(\xi)\sqrt{1+\xi} \right] ,
\label{partprodeqs}
\end{equation}
respectively, for five parameters:  the four-velocity $u^\mu$, the transverse temperature $\Lambda$, and the anisotropy parameter $\xi$.\footnote{Note that the four-velocity satisfies $u^\mu u_\mu =1$ and hence it contains only three independent degrees of freedom.} In the above equations, we use a $\perp$ subscript to indicate two-dimensional vectors in the transverse plane, e.g. ${\bf u_\perp} \equiv (u_x,u_y)$ and ${\bf\nabla_\perp} \equiv (\partial_x,\partial_y)$.   We have also introduced a compact notation for the convective derivative $D_u \equiv u^\mu \partial_\mu$, the longitudinal derivative $D_z \equiv z^\mu \partial_\mu$, and the expansion scalars $\theta_u \equiv \partial_\mu u^\mu$ and $\theta_z \equiv \partial_\mu z^\mu$.

In the most general case, where the matter expands in the longitudinal and transverse directions without any symmetry constraints, one can use the following parametrization of the LAB frame four-velocity of the fluid $u^\mu$ and the space-like four-vector $z^\mu$
\begin{eqnarray}
u^\mu &=& (u_0 \cosh \vartheta, {\bf u_\perp}, u_0 \sinh \vartheta) \,  , \label{U3+1} \\
z^\mu &=& (	 \sinh \vartheta, {\bf 0},  \cosh \vartheta) \, , \label{V3+1}
\end{eqnarray}
where we introduced the longitudinal rapidity of the fluid cell $\vartheta$. Using the four-velocity normalization condition, $u^\mu u_\mu =1$, one has
\begin{eqnarray}
u_0 &=& \sqrt{1+u_\perp^2} \, , \nonumber \\
u_\perp &\equiv& \sqrt{u_x^2 + u_y^2} \, .
\label{u0}
\end{eqnarray}
With the parametrizations (\ref{U3+1}) and (\ref{V3+1}), one may calculate the following quantities appearing in Eqs.~(\ref{enmomconU})-(\ref{partprodeqs}),
\begin{eqnarray}
D_u &=& {\bf u}_\perp \cdot {\bf \nabla}_\perp + u_0 \hat{L}_1 \, , \\
\theta_u &=& {\bf \nabla}_\perp \cdot {\bf u}_\perp + \hat{L}_1 u_0 + u_0 \hat{L}_2 \vartheta \, , \\
D_z &=&  \hat{L}_2 \, , \\
\theta_z &=&  \hat{L}_1 \vartheta \, , \\
u_\nu D_z z^\nu &=& u_0 \hat{L}_2 \vartheta \, , \\
z_\nu D_u u^\nu &=& - u_0 \left( {\bf u}_\perp \cdot 
{\bf \nabla}_\perp + u_0 \hat{L}_1  \right) \vartheta \, ,
\label{op1}
\end{eqnarray}
where the two linear differential operators, $\hat{L}_1$ and $\hat{L}_2$, are given by
\begin{eqnarray}
\hat{L}_1  &=& \cosh (\varsigma - \vartheta) \partial_\tau - \sinh (\varsigma - \vartheta) \frac{\partial_\varsigma}{\tau}, \label{op21}\\
-\hat{L}_2  &=& \sinh (\varsigma - \vartheta) \partial_\tau - \cosh (\varsigma - \vartheta) \frac{\partial_\varsigma}{\tau}.\label{op22}
\end{eqnarray}
We also use the relation between the relaxation time $\tau_{\rm eq}$ and the shear viscosity to entropy density ratio $\bar{\eta} \equiv \eta /s$ \cite{Martinez:2010sc},\footnote{We note that the factor of 2 in the denominator of Eq.~(\ref{reltime}) is needed if one uses the spheroidal form (\ref{distansatz}) together with the zeroth and first moments of the Boltzmann equation.}
\begin{eqnarray}
\tau_{\rm eq} = \frac{5 \bar{\eta}}{2 T}.
\label{reltime}
\end{eqnarray}

\subsection{Anisotropic equation of state}
\label{ssec:eos}

Herein, we consider a system that consists of massless particles described by the anisotropic distribution function (\ref{distansatz}). Using standard kinetic theory definitions
\begin{eqnarray}
N^\mu &\equiv& \int d^3P \,p^\mu f , \\
T^{\mu\nu} &\equiv& \int d^3P \,p^\mu p^\nu f ,
\end{eqnarray}
where $d^3P \equiv d^3p / \left[ (2 \pi)^3 p^0 \right]$, and the tensor decompositions specified in Eqs.~(\ref{Taniso}) and (\ref{Naniso}), one can calculate the thermodynamic properties of the system
\begin{eqnarray}
\label{densaniso}
n(\Lambda, \xi) &=& \frac{n_{\rm iso}(\Lambda)}{\sqrt{1+\xi}}\, ,\\
\label{energyaniso}
{\cal \varepsilon}(\Lambda, \xi) &=& {\cal R}(\xi)\,\varepsilon_{\rm iso}(\Lambda)\, ,\\
\label{transpressaniso}
P_\perp(\Lambda, \xi) &=& {\cal R}_\perp(\xi)\,P_{\rm iso}(\Lambda)\, ,\\
\label{longpressaniso}
P_\parallel(\Lambda, \xi) &=& {\cal R}_\parallel(\xi)\,P_{\rm iso}(\Lambda)\, ,
\end{eqnarray}
where $n_{\rm iso}$, $\varepsilon_{\rm iso}$, and $P_{\rm iso}$ are the isotropic particle density, energy density, and pressure, respectively, and
\begin{eqnarray}
\label{R}
{\cal R}(\xi) &\equiv& \frac{1}{2}\left[\frac{1}{1+\xi}
+\frac{\tan^{-1}\sqrt{\xi}}{\sqrt{\xi}} \right] ,  \\ 
\label{RT}
{\cal R}_\perp(\xi) &\equiv& \frac{3}{2 \xi} 
\left[ \frac{1+(\xi^2-1){\cal R}(\xi)}{\xi + 1}\right]
 \, , 
\\ 
\label{RL}
{\cal R}_\parallel(\xi) &\equiv&  \frac{3}{\xi} 
\left[ \frac{(\xi+1){\cal R}(\xi)-1}{\xi+1}\right] .
\end{eqnarray}
Herein, we assume the simple case of a conformal fluid, i.e. $\varepsilon_{\rm iso} = 3 P_{\rm iso}$.  As a result, Eqs.~({\ref{densaniso}})--({\ref{longpressaniso}}) describe the equation of state of an anisotropic system of classical massless particles with vanishing chemical potential.
%

\subsection{Initial conditions}
\label{ssec:ini}

In order to solve the set of partial differential equations (\ref{enmomconU})--(\ref{partprodeqs}) in general (non-boost-invariant (3+1)-dimensional evolution), one has to make a reasonable assumption about the initial conditions at the initial longitudinal proper-time for the hydrodynamic evolution, $\tau = \tau_0$, i.e. one has to define five three-dimensional profiles: $\Lambda (\tau_0, {\bf x_\perp}, \varsigma)$, $\xi (\tau_0, {\bf x_\perp}, \varsigma)$, $u_x (\tau_0, {\bf x_\perp}, \varsigma)$, $u_y (\tau_0, {\bf x_\perp}, \varsigma)$, and  $\vartheta (\tau_0, {\bf x_\perp}, \varsigma)$. 

During a heavy-ion collision, due to inelastic interactions the participating nucleons deposit some energy in the space-time volume of the fireball. In this work, we assume that the distribution of deposited energy 
is well described by the optical Glauber model.\footnote{Although it is quite interesting, for this first study we do not take into account initial fluctuations in the position of the nucleons or nucleonic substructure. We postpone the Monte-Carlo event-by-event analysis to a future work.} Herein, we assume that the initial energy density is proportional to the scaled initial density of the sources. Therefore, the transverse momentum scale is given by 
\begin{eqnarray}
\Lambda (\tau_0, {\bf x_\perp}, \varsigma) = \varepsilon^{-1}_{\rm iso}\!\left( \varepsilon_0 \frac{\rho(b, {\bf x_\perp}, \varsigma)}{\rho(0, {\bf 0}, 0)} \right) ,
\label{inilambda}
\end{eqnarray}
where the proportionality constant $\varepsilon_0$ is chosen in such a way as to reproduce the total number of charged particles measured in the experiment, and $\varepsilon^{-1}_{\rm iso}$ denotes the inverse $\varepsilon_{\rm iso}(\Lambda)$ function.\footnote{In principle, one could use the full expression for the energy density given by Eq. (\ref{energyaniso}) in Eq. (\ref{inilambda}), however, since we use an initial anisotropy profile that is homogeneous in space, this would merely result in the overall multiplicative factor which can be absorbed by rescaling $\varepsilon_0$.}

The density of sources is constructed using the following mixed model
\begin{eqnarray}
\rho(b, {\bf x_\perp}, \varsigma) &\equiv & \left[\!\frac{}{} (1 - \kappa) (\rho_{\rm WN}^+(b, {\bf x_\perp}) +\rho_{\rm WN}^-(b, {\bf x_\perp})) \right. \nonumber \\
& & \left. + \, 2 \,\kappa\, \rho_{\rm BC}(b, {\bf x_\perp}) \frac{}{}\!\right] f(\varsigma - \varsigma_S({b, \bf x_\perp})) \, ,
\label{sources}
\end{eqnarray}
where $\rho^{\pm}_{\rm WN}$ is the density of wounded nucleons from the left/right-moving nuclei and $\rho_{\rm BC}$ is the density of binary collisions, both of which are obtained using the optical limit of the Glauber model
\begin{eqnarray}
\rho_{\rm WN}^\pm(b, {\bf x_\perp})  &\equiv& T\left( {\bf x_\perp}\!\mp\!\frac{{\bf b_\perp}}{2}\right)\!\left[ 1\! -\! e^{- \sigma_{in} T\left( {\bf x_\perp} \pm  \frac{{\bf b_\perp}}{2}\right)} \right], \hspace{5mm} \label{sourcesGlauber1}\\ 
\rho_{\rm BC}(b, {\bf x_\perp})  &\equiv& \sigma_{in}T\left( {\bf x_\perp}\!+\! \frac{{\bf b_\perp}}{2}\right) T\left( {\bf x_\perp}\!-\!\frac{{\bf b_\perp}}{2}\right) .
\label{sourcesGlauber2}
\end{eqnarray}
The longitudinal profile is taken to be
\begin{eqnarray}
f(\varsigma) \equiv \exp \left[ - \frac{(\varsigma - \Delta \varsigma)^2}{2 \sigma_\varsigma^2} \Theta (|\varsigma| - \Delta \varsigma) \right] .
\label{longprof}
\end{eqnarray}

For the LHC case studied here, we use $\kappa = 0.145$ for the mixing factor and an inelastic cross-section of $\sigma_{ in} = 62$ mb. We also restrict ourselves to the minimum-bias studies in which case $b \equiv |{\bf b}| =  9.5$ fm. The parameters of the longitudinal profile (\ref{longprof}) were fitted to reproduce the pseudorapidity distribution of charged particles with the results being $\Delta\varsigma = 2.5$ and $\sigma_{\varsigma} = 1.4$. The shift in rapidity is calculated according to the formula \cite{Bozek:2009ty}
\begin{eqnarray}
\varsigma_S \equiv \frac{1}{2}\ln \frac{\rho_{\rm WN}^++ \rho_{\rm WN}^- + v_P (\rho_{\rm WN}^+- \rho_{\rm WN}^-)}{\rho_{\rm WN}^++ \rho_{\rm WN}^- - v_P (\rho_{\rm WN}^+- \rho_{\rm WN}^-)} \, ,
\label{shift}
\end{eqnarray}
where all functions are understood to be evaluated at a particular value of $b$ and ${\bf x_\perp}$. The participant velocity is defined as $v_P \equiv \sqrt{(\sqrt{s}/2)^2-(m_N/2)^2}/(\sqrt{s}/2)$ and $m_N$ is the nucleon mass. In Eqs.~(\ref{sourcesGlauber1})--(\ref{sourcesGlauber2}) we have made use of the thickness function
\begin{eqnarray}
T({\bf x_\perp}) \equiv \int dz \,\rho_{\rm WS}({\bf x_\perp},z) \, ,
\label{thickness}
\end{eqnarray}
where the nuclear density is given by the Woods-Saxon profile
\begin{equation}
\rho_{\rm WS}({\bf x_\perp},z) \equiv \rho_0 \left[ 1 + \exp\left(\frac{\sqrt{{\bf x_\perp}^2 + z^2} - R}{a}\right)\right]^{-1}.
\label{WS}
\end{equation}
For Pb-Pb collisions, we use $\rho_0 = 0.17\, {\rm fm}^{-3}$ for the nuclear saturation density, $R = 6.48$ fm for the nuclear radius, and $a = 0.535$ fm for the surface diffuseness of the nucleus. 

In the calculations presented herein, we assumed that the produced matter has initially no transverse flow, i.e. $u_x (\tau_0, {\bf x_\perp}, \varsigma) = u_x (\tau_0, {\bf x_\perp}, \varsigma) = 0$, while the initial longitudinal flow is of Bjorken form $\vartheta (\tau_0, {\bf x_\perp}, \varsigma) = \varsigma$. For simplicity, the initial anisotropy parameter is assumed to be homogeneous, $\xi (\tau_0, {\bf x_\perp}, \varsigma) = \xi_0$.\footnote{On general grounds, one can expect that the level of momentum-space anisotropy is larger in regions that have a lower effective temperature.  As a result, our assumption of a constant $\xi_0$ is a conservative one.}
%
\begin{figure*}[t]
\centerline{
\hspace{7mm}
\includegraphics[width=0.56\linewidth]{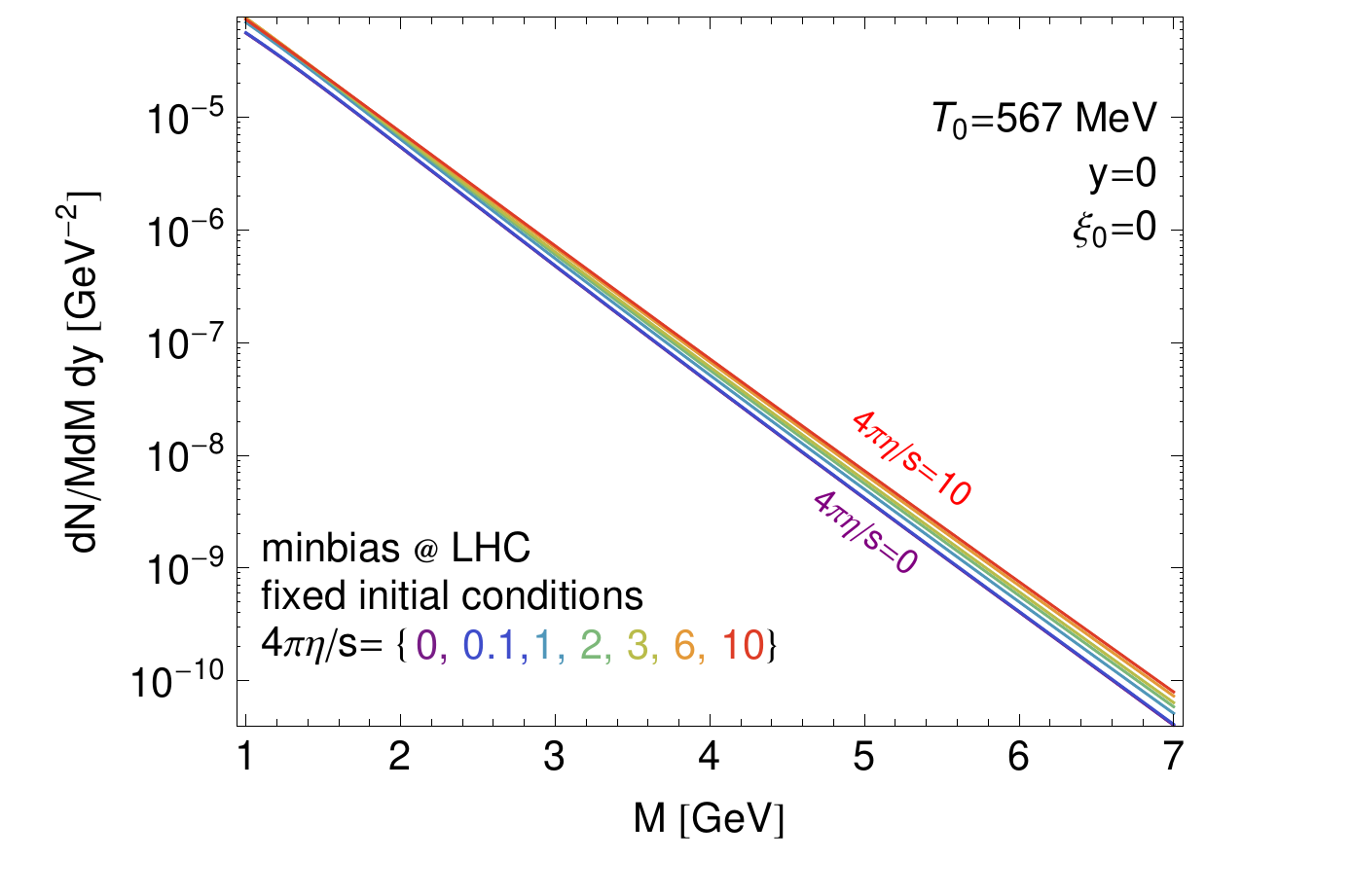}
\hspace{-1.2cm}
\includegraphics[width=0.56\linewidth]{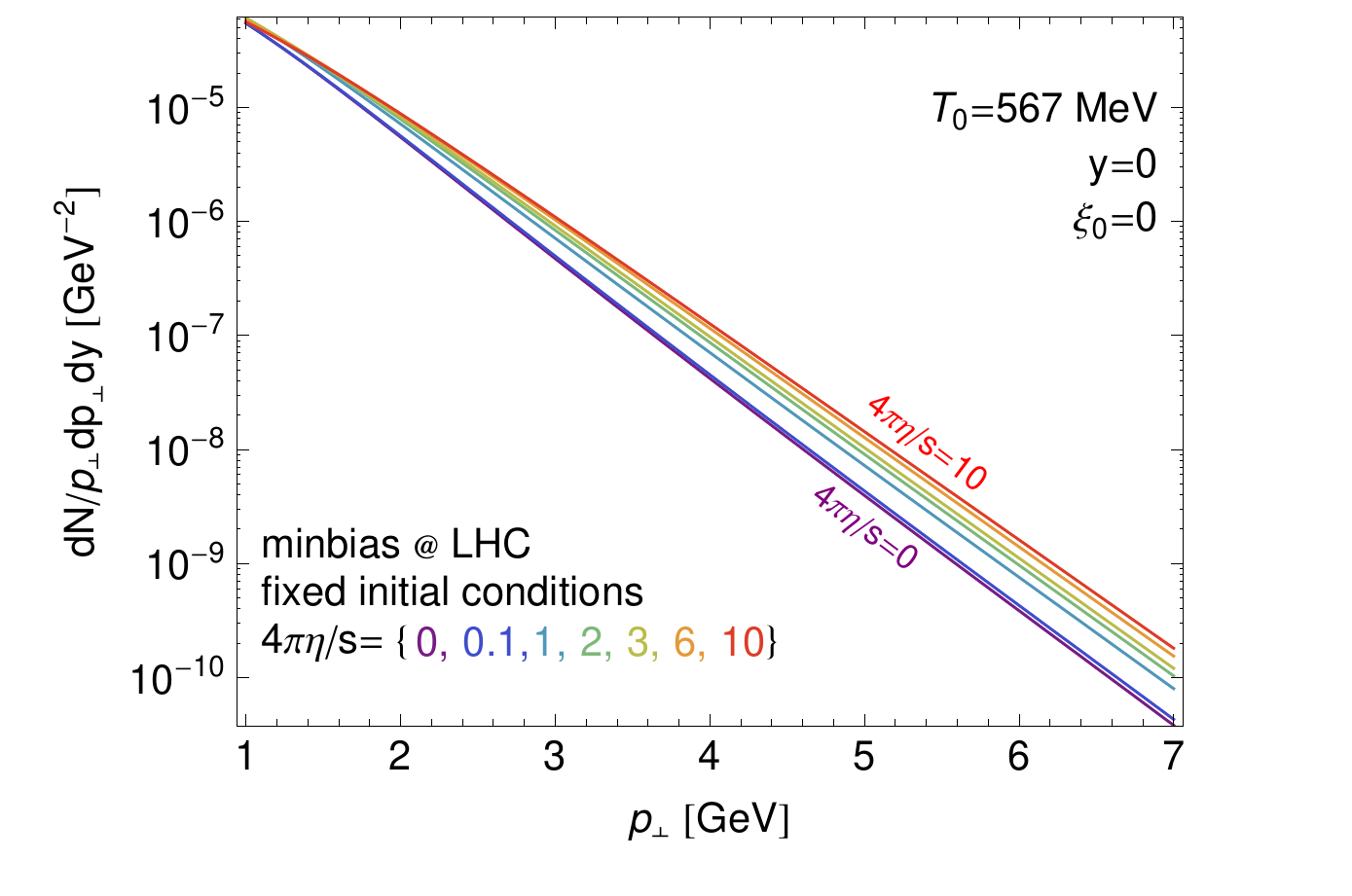}
}
\vspace{-3mm}
\caption{
(Color online) The invariant mass spectra (left) and transverse momentum spectra (right) of dilepton pairs at
midrapidity, $y = 0$, for various values of shear viscosity to entropy density ratio $4 \pi {\bar \eta} \in \{0, 0.1, 1, 2, 3, 6, 10\}$. For all cases the initial
temperature is fixed to $T_0 = 567$ MeV and the system is initially isotropic in momentum space, $\xi_0 = 0$.
}
\label{spectrafixini}
\end{figure*}
%
\begin{figure*}[t]
\centerline{
\hspace{7mm}
\includegraphics[width=0.56\linewidth]{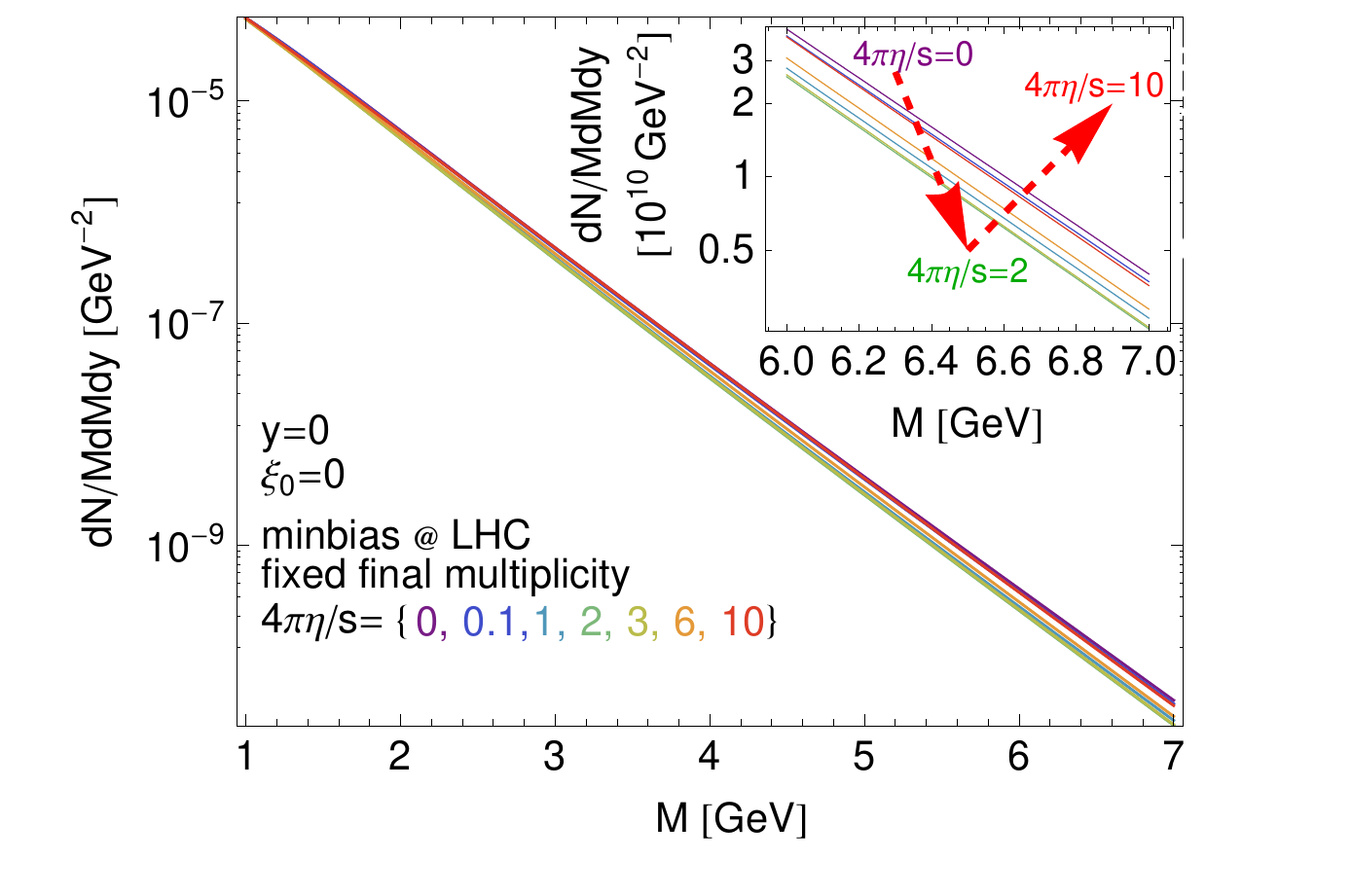}
\hspace{-1.2cm}
\includegraphics[width=0.56\linewidth]{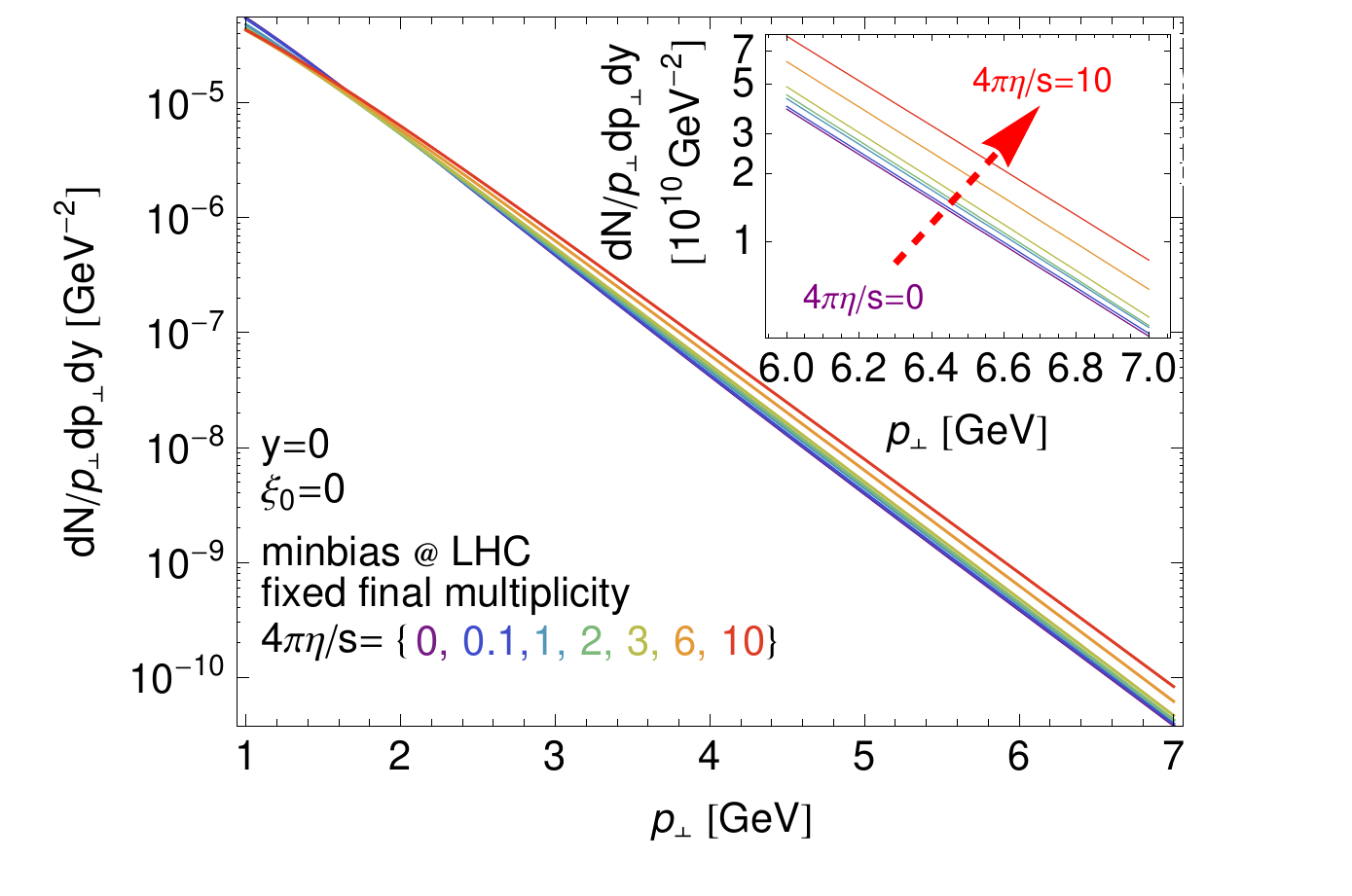}
}
\vspace{-3mm}
\caption{
(Color online) Same as Fig.~\ref{spectrafixini}, except here, instead of fixing the initial temperature, we keep the final particle multiplicity fixed.
}
\label{spectrafixfinal}
\end{figure*}

\section{Results}
\label{sec:results}

In this section, we present our model predictions for the minimum-bias $e^+ e^-$ yields resulting from Pb-Pb collisions at LHC with $\sqrt{s}=2.76$ TeV beam energy. Before presenting our results, we first explain the setup and parameters chosen for our calculations.

Since the differential dilepton rate $d R^{l^+l^-}\!/d^4\!P$ given in Eq.~(\ref{kineticratefinal}) is independent of the assumed space-time model, we first evaluate it numerically using double-exponential integration on a uniformly-spaced 4-dimensional grid in $M/\Lambda$, $p_\perp/\Lambda$, $y$, and $\log_{\rm 10} (\xi +1)$ such that $M/\Lambda, p_\perp/\Lambda \in \{0.1, 40\}$, $y \in \{-6,6\}$ and $\log_{\rm 10} (\xi + 1) \in \{-1, 3\}$.\footnote{Note that the dilepton rate is an even function of $y$, therefore, in practice, we may restrict ourself to positive values of $y$ only.} The spacing was chosen in such a way that, after building a four-dimensional interpolating function from the table, we could assume that it is valid at continuous values of these four variables. We then evaluated the remaining integrations over space-time, transverse momentum angle, and transverse momentum or invariant mass appearing in Eqs.~(\ref{spectrumeqs}) using Monte Carlo integration.  For the integration over the transverse momentum we have specified the default cuts as follows: $p_\perp^{\rm min} =1$ GeV and $p_\perp^{\rm max} = 20$ GeV, while for the invariant mass integration we used $M^{\rm min} = 1$ GeV and $ M^{\rm max} = 20$ GeV.

\subsection{Dilepton production with fixed initial conditions}
\label{ssec:fixini}

We begin by presenting the minimum-bias dilepton spectra at midrapidity, $y = 0$, calculated assuming fixed initial conditions. The initial central temperature was taken to be $T_0 = 567$ MeV, at a starting time of $\tau_0 = 0.3$ fm/c. For this case, we fixed
the initial anisotropy parameter to $\xi_0 = 0$, which means that the system is initially isotropic in momentum space. In Fig.~\ref{spectrafixini} we plot the resulting invariant mass spectra (left) and transverse momentum spectra (right) of
dilepton pairs for various values of $\bar{\eta}$. We can see that the spectra flatten with increasing $\bar{\eta}$ and the normalization increases with increasing $\bar{\eta}$. The latter implies that the total final multiplicity changes with changing $\bar{\eta}$. This effect is more visible in the $p_\perp$ spectra. We note here that the fixed-initial-condition behavior described above is monotonic in $\bar{\eta}$.  The increase in multiplicity with increasing $\bar\eta$ is related to dissipative particle production in the QGP, which would also be reflected in increased final particle multiplicity across all particle types. 
%
\begin{figure*}[t]
\centerline{
\hspace{7mm}
\includegraphics[width=0.56\linewidth]{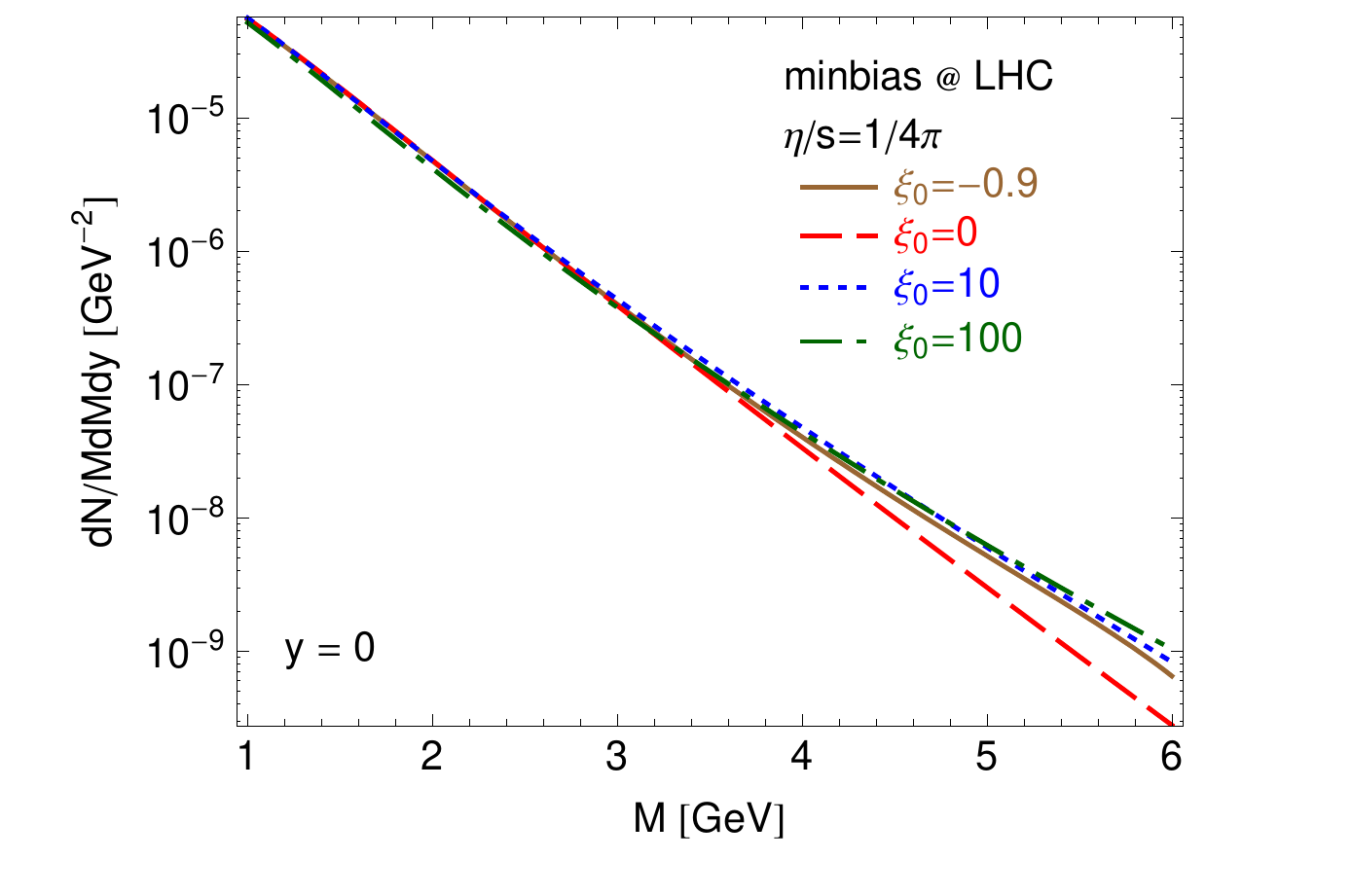}
\hspace{-1.2cm}
\includegraphics[width=0.56\linewidth]{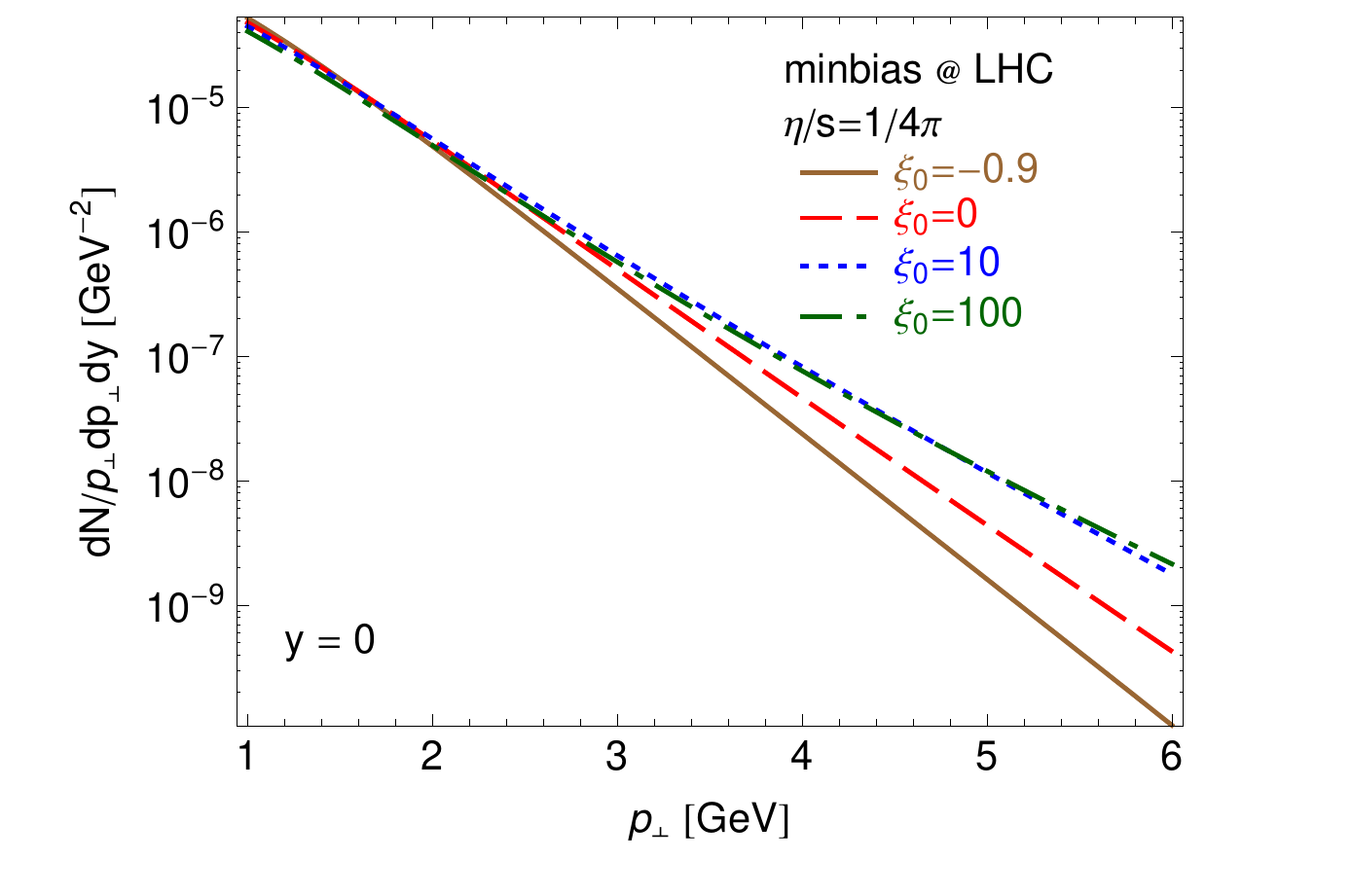}
}
\vspace{-3mm}
\caption{
(Color online) The invariant mass spectra (left) and transverse momentum
spectra (right) of dilepton pairs at midrapidity, $y = 0$, for various initial anisotropy conditions, and $4 \pi \bar \eta = 1$. The results with
$\xi_0 = -0.9, 0, 10$, and $100$ are denoted by brown solid, red dashed, blue dotted and green dot-dashed lines, respectively.
}
\label{spectravarxi1}
\end{figure*}
%
%
\begin{figure*}[t]
\centerline{
\hspace{7mm}
\includegraphics[width=0.56\linewidth]{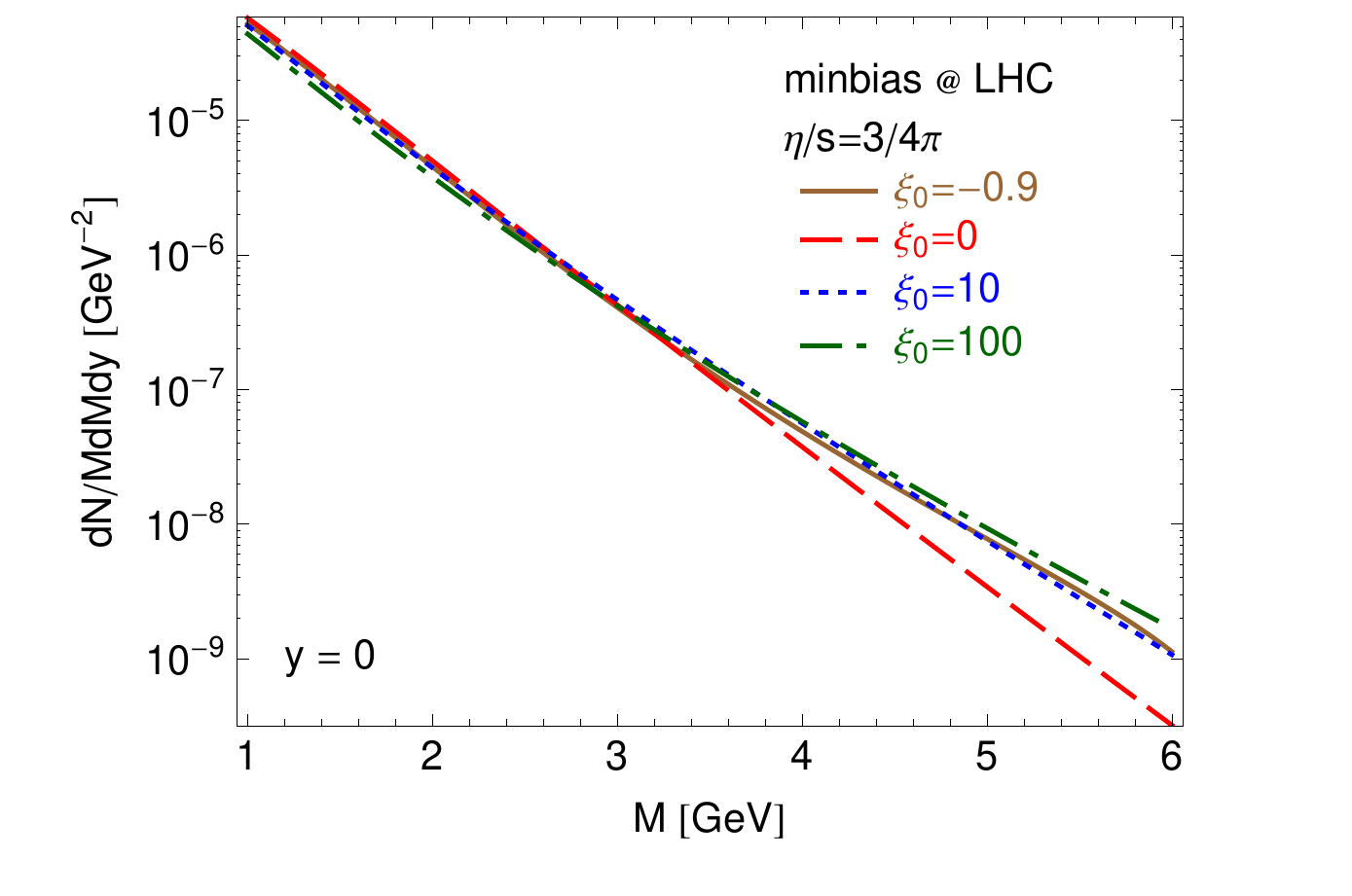}
\hspace{-1.2cm}
\includegraphics[width=0.56\linewidth]{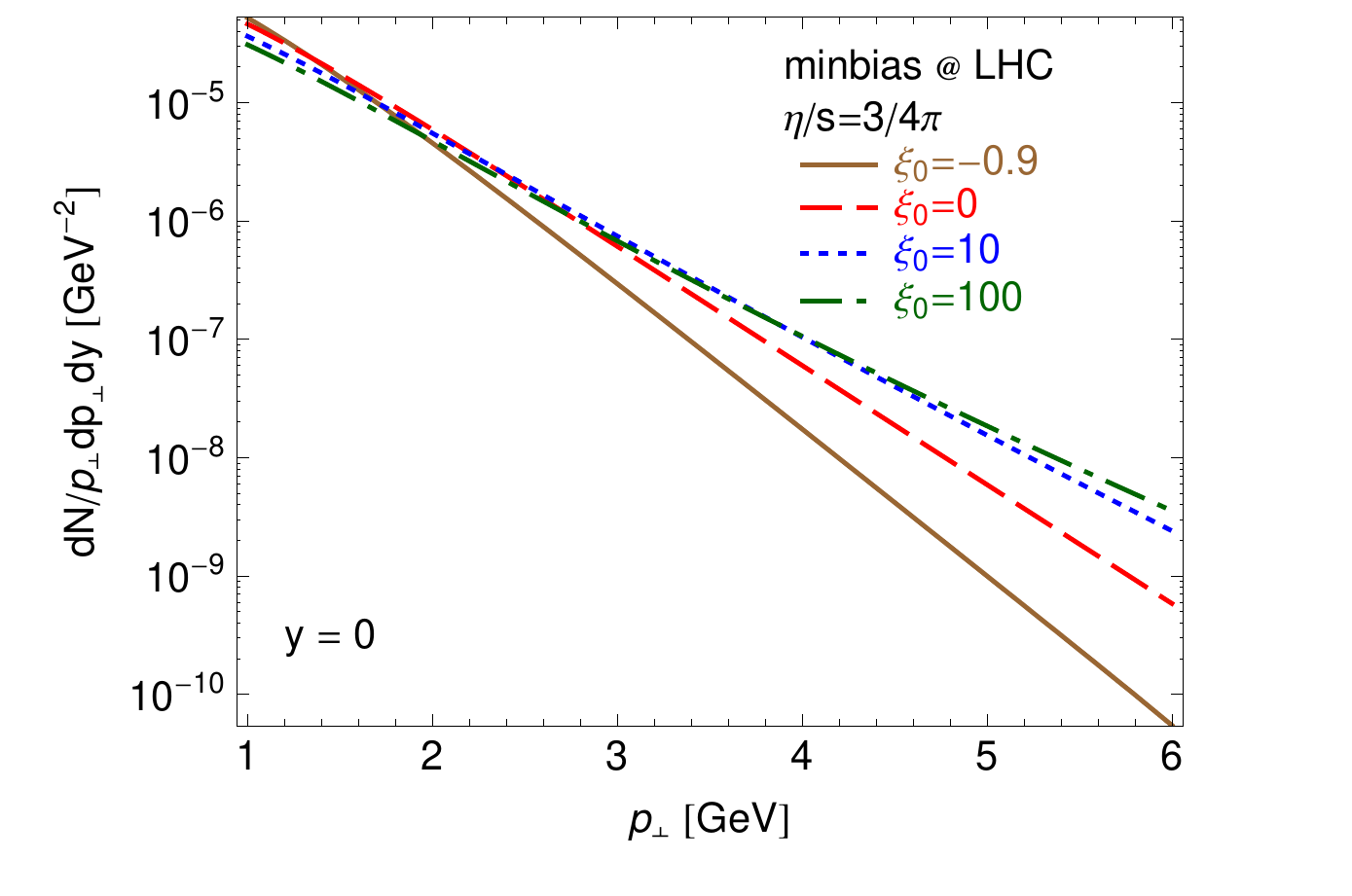}
}
\vspace{-3mm}
\caption{
(Color online) Same as Fig.~\ref{spectravarxi1} however here we take  $4 \pi \bar \eta = 3$.
}
\label{spectravarxi2}
\end{figure*}

\subsection{Dilepton production with fixed final multiplicity}
\label{ssec:fixfin}

The case presented in Section \ref{ssec:fixini} is unphysical since the average final particle multiplicity in a given centrality range is held fixed when presenting experimental results for the dilepton spectra.  We presented the prior case only to establish that, for fixed initial temperature, the behavior seen in the final dilepton spectra is monotonic when $\bar\eta$ is increased.  In this section, we present the same analysis, however, now, for each presented case, the initial central temperature is rescaled in such a way as to keep the final multiplicity of particles at freeze-out fixed.

Fig.~\ref{spectrafixfinal} presents the invariant mass spectra (left) and transverse momentum spectra (right) of dilepton pairs at midrapidity, $y = 0$. In the left panel, we can clearly see the effect of the rescaling of the initial temperature
for cases with various $\bar\eta$, i.e. the spectra does not change significantly as long as the final multiplicity of particles at freeze-out is fixed. However, importantly, we observe that the spectra do not necessarily have a monotonic dependence as $\bar\eta$ is increased (see e.g. the left panel of Fig.~\ref{spectrafixini}). 

The non-monotonic behavior primarily due to the fact that particle production within anisotropic hydrodynamics is not a monotonic function of $\bar \eta$, as it is for standard viscous hydrodynamics \cite{Martinez:2012tu,Florkowski:2013lza,Bazow:2013ifa}. Instead, one observes a maximum in particle production at a certain value of $\bar\eta$ which depends on the assumed initial temperature.  The fact that there must be a maximum can be anticipated by the fact that particle production should vanish in the both the ideal and free streaming limits.\footnote{The dependence of particle production on the assumed value of $\bar\eta$ is discussed in more detail in Appendix \ref{sec:app1}.  In that appendix, we compare particle production as a function of $\bar\eta$ using both anisotropic and viscous hydrodynamics.}  As a result, when fixing the initial temperature to guarantee fixed final multiplicity, the required temperature may not be monotonically decreasing as $\bar\eta$ is increased.

Note that, although the final multiplicity of particles created at freeze-out is fixed, the number of dileptons which are produced in the QGP volume varies with $\bar\eta$. As a result, we observe a small but noticeable non-monotonic change in the dilepton invariant mass spectra.  Similar arguments also apply to the transverse momentum spectra shown in the right panel of Fig.~\ref{spectrafixfinal}.  For the $p_\perp$-spectra the effect is smaller.  Based on our final results shown in Fig.~\ref{spectrafixfinal} one can see that for $4 \pi {\bar \eta} \in (1, 3)$, which spans the range of ${\bar \eta}$ extracted from the flow experimental data, the impact of shear viscosity in the system on the dilepton spectra is quite small.  

\subsection{Effect of initial anisotropy}
\label{ssec:inianiso}

We now turn to the analysis of the impact of the initial anisotropy in the system, $\xi_0$, on the dilepton spectra. In Figs.~\ref{spectravarxi1} and ~\ref{spectravarxi2} we present invariant mass spectra (left panels) and transverse momentum spectra (right panels) for dilepton pairs at midrapidity, $y = 0$, for various initial anisotropy conditions.  The results with $\xi_0 = -0.9, 0, 10$, and $100$ are denoted by brown solid, red dashed, blue dotted and green dot-dashed lines, respectively. The values of $\xi_0 < 0$ ($\xi_0 > 0$) correspond to prolate (oblate) initial momentum distribution functions. In Fig.~\ref{spectravarxi1} we keep the viscosity fixed to ${\bar \eta}  = 1/4 \pi$ while in Fig.~\ref{spectravarxi2} we set ${\bar \eta}  = 3/4 \pi$. In each case the initial energy density at the center, $\varepsilon_0$, is rescaled to keep the final multiplicity of particles at freeze-out fixed.  The values of $\varepsilon_0$ used in each case are listed in Table \ref{table:edensities}.
\begin{table}[t]
  \begin{center}
    \begin{small}
      \begin{tabular}{|c|cccc|}
      \hline
\diaghead{\theadfont xcgfxgdfsshjfdn}%
{$\quad 4 \pi \eta/s$}{$\xi_0\quad$} & -0.9 & 0 & 10 & 100 \\   \hline 0.1 & - & 72.11 & -  & - \\ 
\hline 1 & 12.98 &  64.53 & 235.86 & 714.31 \\
\hline 3 & 13.69 & 60.44 & 215.61 &  660.74 \\
\hline
      \end{tabular}
    \end{small}
  \end{center}
  \caption{\small Values of the initial central energy density, $\varepsilon_{\rm 0} \,\rm [GeV/fm^3]$, used in all the figures of this Section except for Figs.~\ref{spectrafixini} and ~\ref{spectrafixfinal}.
}
\label{table:edensities}
\end{table}
From Figs.~\ref{spectravarxi1} and \ref{spectravarxi2}, one can clearly see that the transverse momentum spectra are quite sensitive to the initial momentum anisotropy in the system. For an initially oblate configuration, they are becoming flatter. The opposite behavior is observed for an initially prolate configuration. This effect is particularly significant for large values of $p_\perp$. This opens possibility to measure initial anisotropy of the plasma by looking at large $p_\perp$ dilepton pairs at LHC. The behavior of the invariant mass spectra, on the other hand, is more difficult to understand since, in this case, both oblate and prolate initial conditions lead to a flattening of the spectra.

\begin{figure}[t]
\includegraphics[scale=0.66]{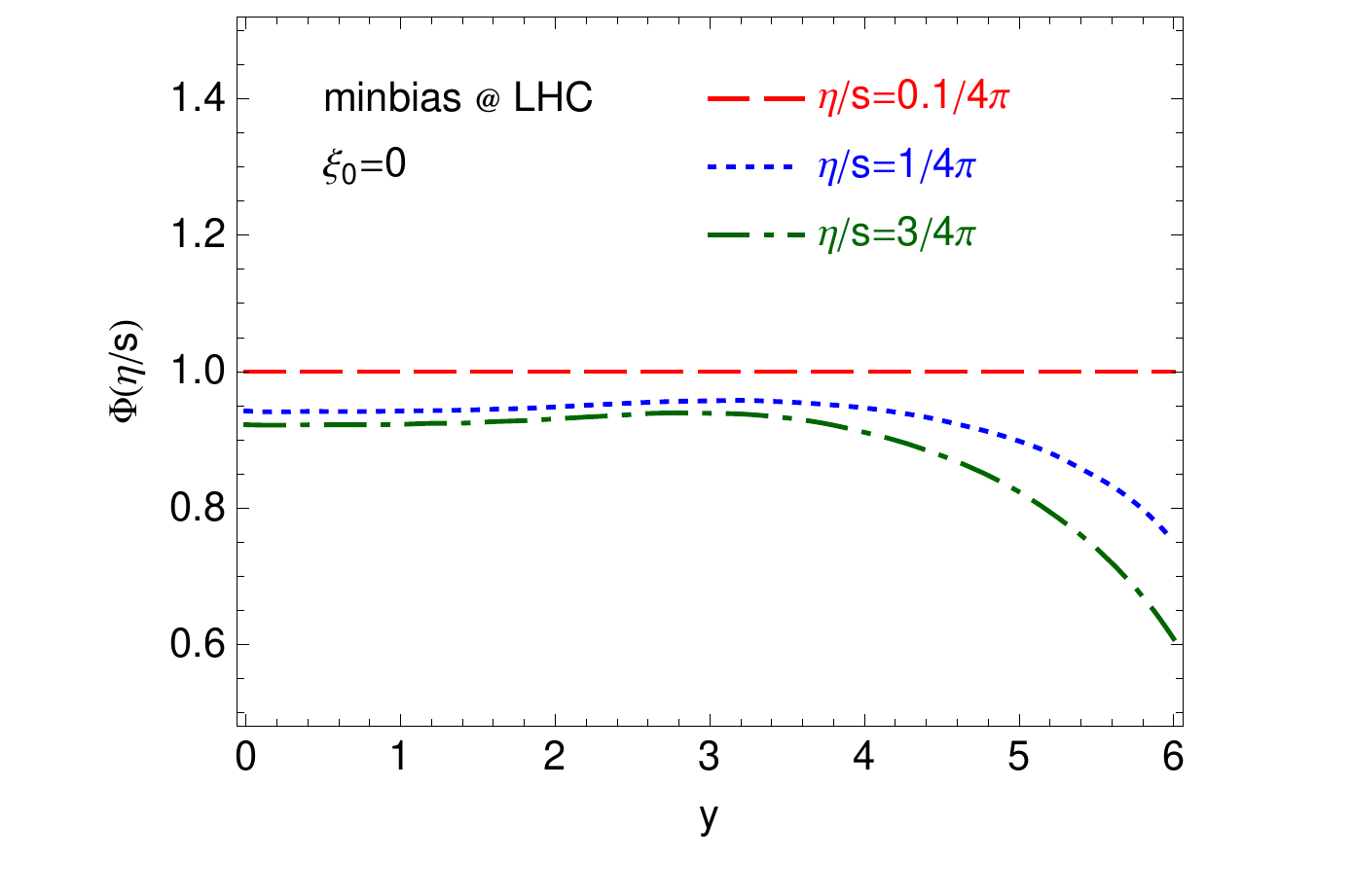}
\caption{
(Color online) The rapidity dependence of the dilepton modification factor $\Phi ({\bar \eta})$ for $4 \pi {\bar \eta} \in \{0.1, 1, 3\}$ denoted by red dashed, blue dotted and green dashed-dotted lines, respectively. The initial anisotropy is $\xi_0 = 0$ in this case. In this case we also use default cuts: $p_\perp^{min} =1$ GeV, $p_\perp^{max} = 20$ GeV, $M^{min} = 1$ GeV and $ M^{max} = 20$ GeV.
}
\label{fig:forward1}
\end{figure}

\begin{figure}[t]
\includegraphics[scale=0.66]{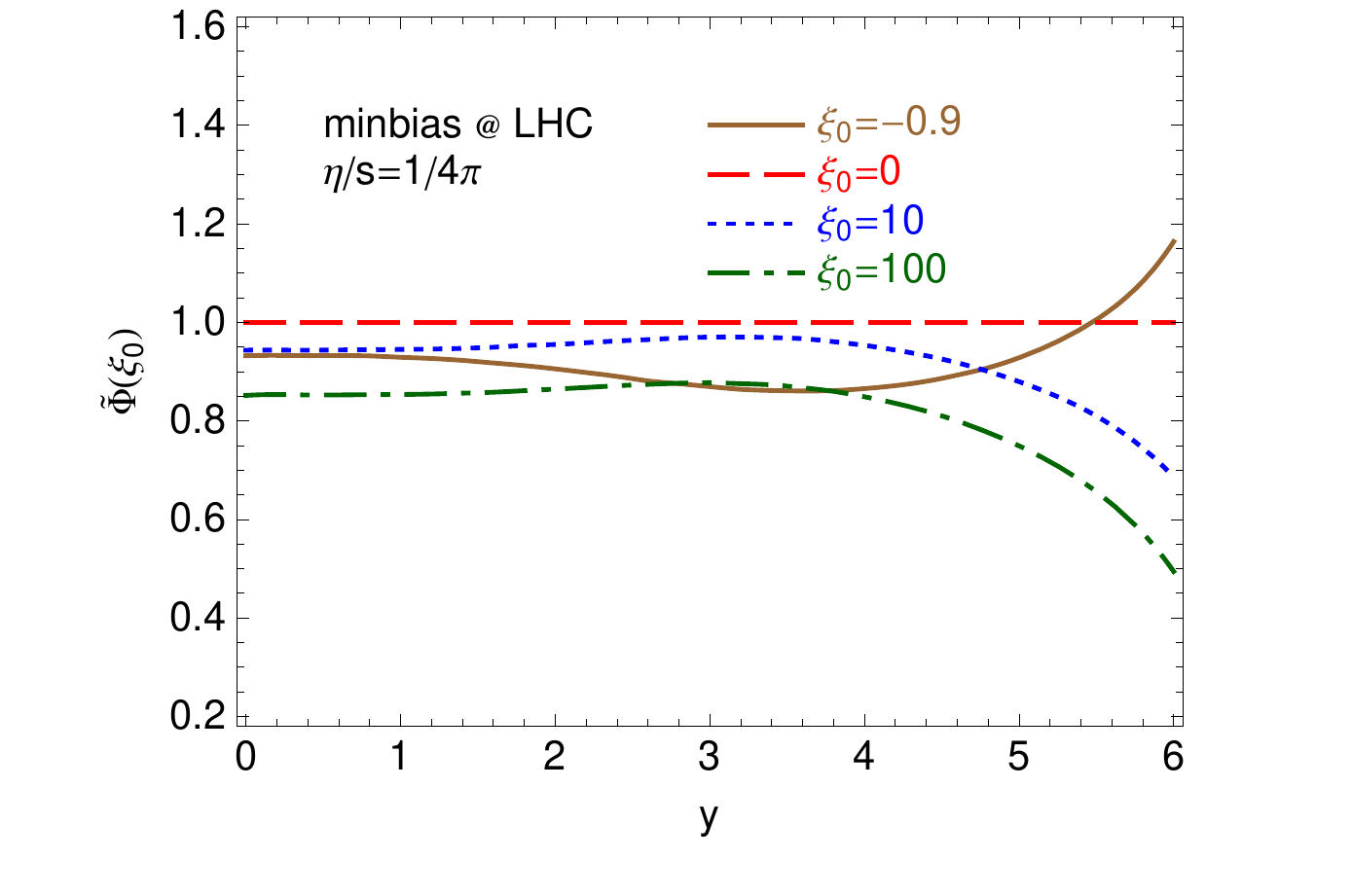}\newline
\includegraphics[scale=0.66]{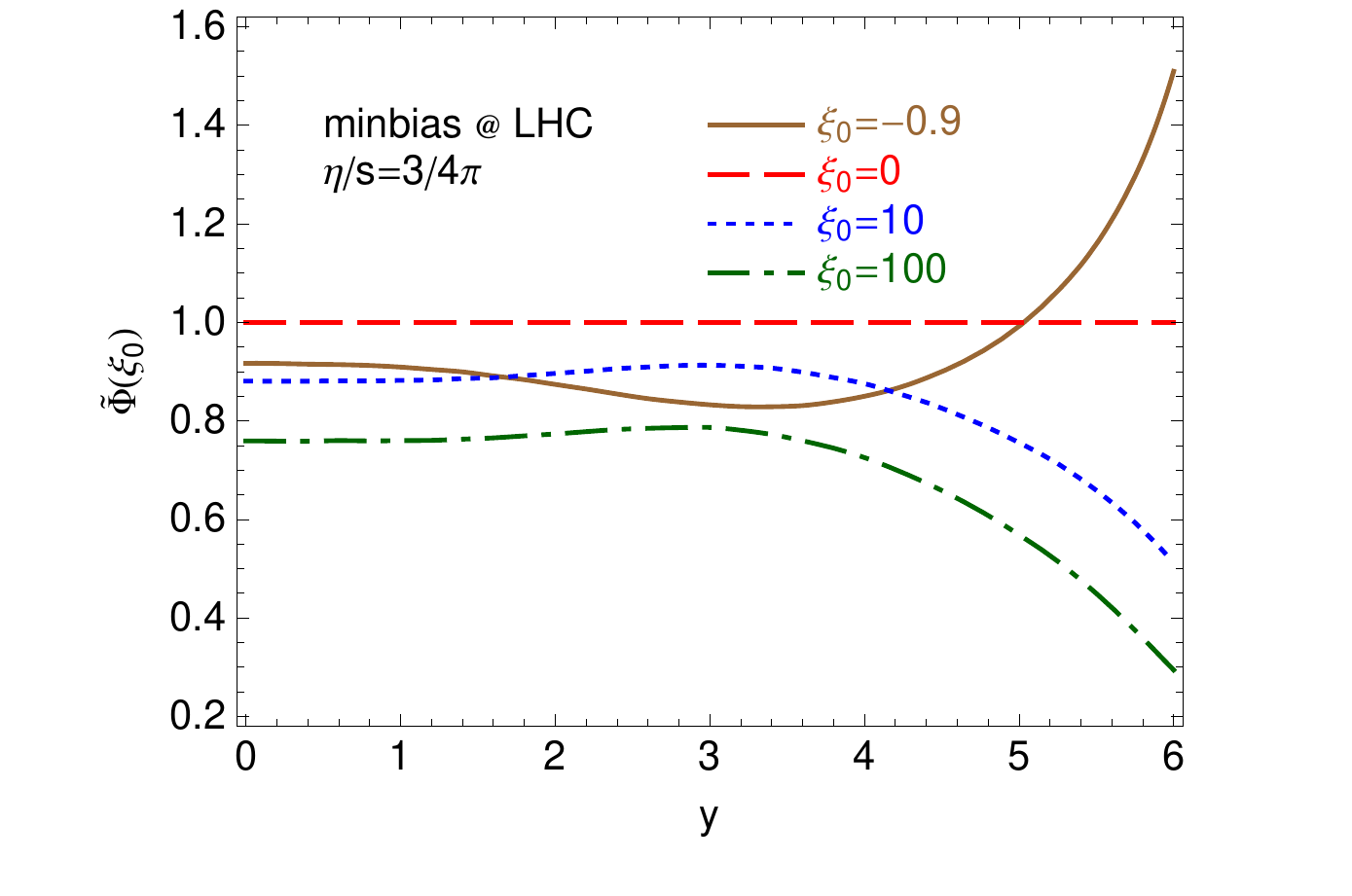}
\caption{
(Color online) The rapidity dependence of the dilepton modification factor ${\tilde \Phi} (\xi_0)$ for $\xi_0 = -0.9, 0, 10$ and $100$ (notation is the same as in Section \ref{ssec:inianiso}) and for $4 \pi {\bar \eta} = 1$ (top panel) and $4 \pi {\bar \eta} = 3$ (bottom panel). The $p_\perp$ and $M$ cuts are the same as in Fig.~\ref{fig:forward1}.
}
\label{fig:forward2}
\end{figure}

\begin{figure}[t]
\includegraphics[scale=0.66]{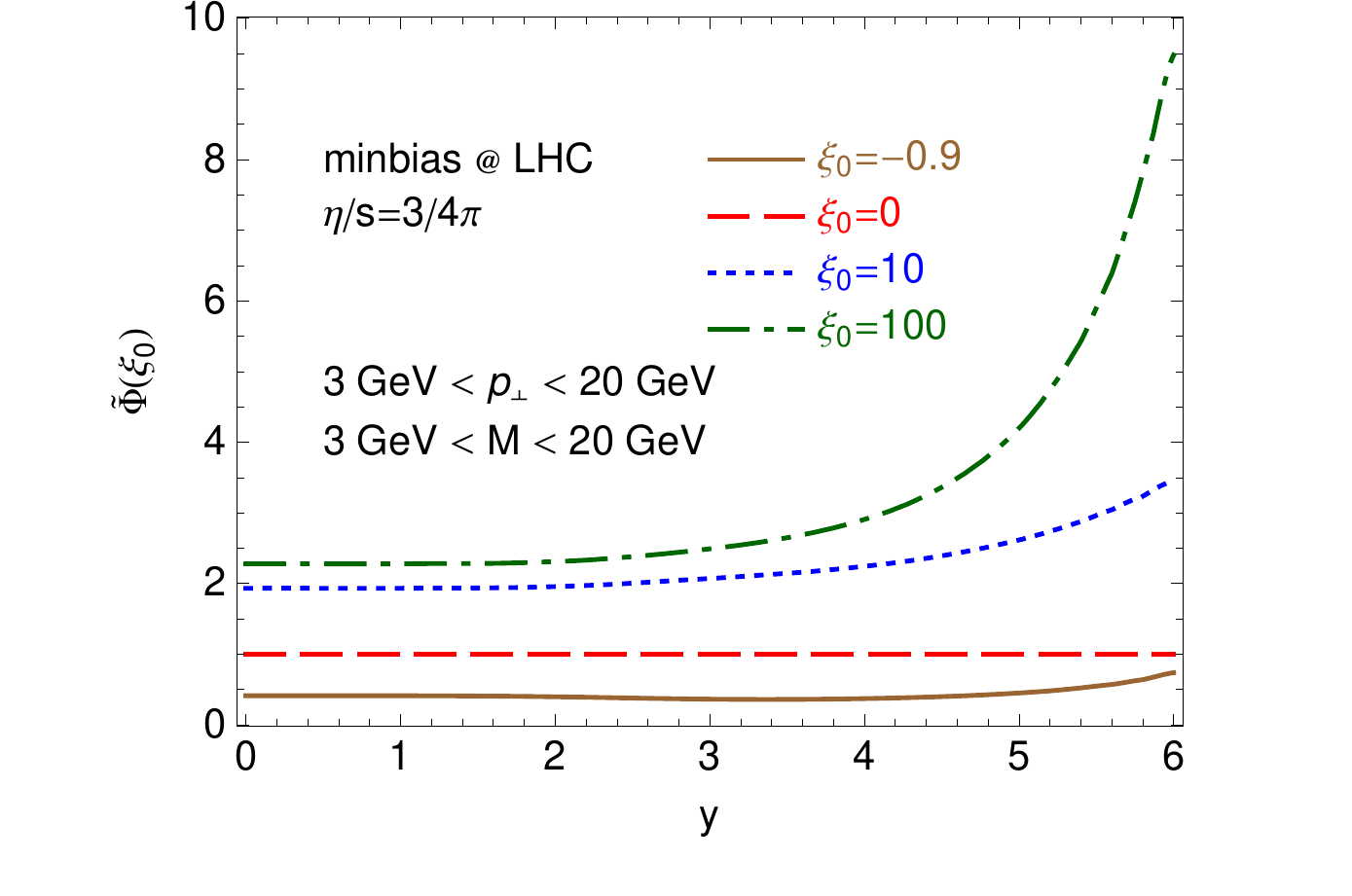}
\caption{
(Color online) Same as the bottom panel of Fig.~\ref{fig:forward2}, except high-energy cuts of 3 GeV $< p_\perp <$ 20 GeV  and 3 GeV $< M <$ 20 GeV.
}
\label{fig:forward3}
\end{figure}

\subsection{Production at forward rapidities}
\label{ssec:forward}
%
We close this Section by presenting an analysis of dilepton production at the forward rapidities following the preliminary study made in Ref.~\cite{Martinez:2008mc}. In Fig.~\ref{fig:forward1} we present the dilepton modification factor 
\begin{equation}
\Phi({\bar \eta}) \equiv \left.\left(\frac{dN^{e^+ e^-} ({\bar \eta})}{dy} \right)\right/ \left(\frac{dN^{e^+ e^-} ({\bar \eta} = 0.1/ 4 \pi)}{dy} \right) ,
\label{mfac1}
\end{equation}
for $4 \pi {\bar \eta} \in \{0.1, 1, 3\}$ and fixed value of $\xi_0 = 0$. In this figure, one sees that increasing the value of ${\bar \eta}$ results in a suppression of particle production. The emission is more suppressed when we go to more forward rapidities, up to 40$\%$ in the most extreme case. In Fig.~\ref{fig:forward2}, we present a complementary study of the dilepton modification factor (analogous to (\ref{mfac1}))
\begin{equation}
{\tilde \Phi}(\xi_0) \equiv \left.\left(\frac{dN^{e^+ e^-} (\xi_0)}{dy} \right)\right/ \left(\frac{dN^{e^+ e^-} (\xi_0 = 0)}{dy} \right) .
\label{mfac2}
\end{equation}
for $\xi_0 = -0.9, 0, 10$ and $100$ (the notation is the same as in Section \ref{ssec:inianiso}) and for $4 \pi {\bar \eta} = 1$ (top panel) and $4 \pi {\bar \eta} = 3$ (bottom panel). Similarly to Fig.~\ref{fig:forward1}, we observe a suppression of the dilepton production at forward rapidities, which increases with increasing initial anisotropy parameter $\xi_0$ and the viscosity in the system. Moreover, we observe the opposite effect when the distribution is initially prolate. In this case, we find dilepton enhancement at forward rapidities together with moderate suppression in midrapidity.  These effects provide the possibility to probe the initial degree of thermalization of the system by looking at forward rapidity emission of dilepton pairs.   Finally, in Fig.~\ref{fig:forward3}, we present the dilepton enhancement factor $\tilde\Phi$ with different cuts corresponding to 3 GeV $< p_\perp <$ 20 GeV  and 3 GeV $< M <$ 20 GeV.  As can be seen from this figure, high-energy dilepton emissions are more sensitive to the level of momentum-space anisotropic in the quark-gluon plasma.  Of course, since statistics are more limited, high-energy dilepton spectra are usually more difficult to measure accurately.

\section{Conclusions}
\label{sec:conc}

In this paper we computed the dilepton invariant mass and transverse momentum spectra produced from the quark-gluon plasma.  To accomplish this, we used the leading-order (3+1)-dimensional anisotropic hydrodynamics equations obtained from the zeroth and first moments of the Boltzmann equation and assumed a conformal (ideal) equation of state.  The anisotropic hydrodynamics equations solved allow for both azimuthal spatial anisotropy and a realistic rapidity profile.  In this paper we considered a fixed (min-bias) impact parameter.  We found that, when adjusting the initial temperature in order to enforce fixed final particle multiplicity, both the dilepton invariant mass and transverse momentum spectra show only a weak dependence on the assumed value of $\eta/s$.  

A similar conclusion was found in an earlier works that used a much more primitive model of the dynamics \cite{Martinez:2008di,Martinez:2008mc}.  With the inclusion of the full (3+1)-dimensional dynamics using anisotropic hydrodynamics, we are now more confident that the dilepton spectra only have a weak dependence on the assumed value of $\eta/s$.  That being said, in these previous works the possibility of a finite initial momentum-space anisotropy $\xi_0$ was not considered.  In this work we found that the high-mass and high-transverse-momentum dilepton spectra are quite sensitive to the initial level of momentum-space anisotropy.  Additionally, we demonstrated that the rapidity dependence of dilepton production is also sensitive to the initial level of momentum-space anisotropy.  These observations offer some hope that one might be able to experimentally determine information about early-time momentum-space anisotropies generated in heavy-ion collisions using dilepton production.

In this work we made a few simplifying assumptions that will be improved in future works.  The first of these is that we only study min-bias collisions.  The magnitude of the effects seen here could depend on centrality in a non-trivial way since in central collisions the plasma lifetime is significantly longer but the level of momentum-space anisotropy developed dynamically in the center of the fireball will be reduced.  We plan to make a systematic study of the centrality dependence of our results in a forthcoming paper.  Another crucial assumption was that we used only the leading order (Born) rate for dilepton production.  It is possible that inclusion of the next-to-leading order rate could significantly modify our conclusions.  Unfortunately, to the best of our knowledge such a calculation only exists for an isotropic quark-gluon plasma \cite{Aurenche:2002wq}.  It would be very interesting to see if these calculations could be extended to the case of an anisotropic quark-gluon plasma.

Looking forward, one should also consider polarized dilepton emission as suggested in Ref.~\cite{Shuryak:2012nf}.  The polarization asymmetry could be quite sensitive to early-time momentum-space anisotropies and possibly also to the assumed value of $\eta/s$.  Finally, we mention that another ideal observable that should be studied further is the emission of real photons.  This has been studied using viscous hydrodynamics in Refs.~\cite{Dion:2011pp,Chatterjee:2011dw,Shen:2013cca} and using simple models of anisotropy evolution in the plasma \cite{Schenke:2006yp,Schenke:2008hw,Bhattacharya:2008up,Bhattacharya:2008mv,Bhattacharya:2009sb,Bhattacharya:2010sq,McLerran:2014hza}.  It is necessary to extend these studies to include the (3+1)-dimensional evolution of the QGP using anisotropic hydrodynamics in order to draw more firm conclusions about the effect of momentum-space anisotropies on photon production.  We also mention that, like dileptons, a difficult, but necessary, step will be to extend the NLO calculation of photon production first obtained in Refs.~\cite{Arnold:2001ba,Arnold:2001ms,Arnold:2002ja} to an anisotropic quark-gluon plasma.  The difficulty in this calculation stems from the presence of color plasma instabilities that render the NLO rate formally infinite.  In practice, these infinities will be regulated due to the eventual saturation of unstable mode growth, but how to implement this in practice is an open question.

\begin{acknowledgments}
R.R. was supported by Polish National Science Center Grant No. DEC-2012/07/D/ST2/02125.
M.S. was supported by U.S. DOE Award No. DE-AC0205CH11231.
\end{acknowledgments}

\appendix

\section{Particle production in viscous and anisotropic hydrodynamics}
\label{sec:app1}

One can show that particle production within anisotropic hydrodynamics is not a monotonically increasing function
of shear viscosity to entropy density, $\eta/s$.  This behavior is in agreement with exact solutions of RTA Boltzmann equation \cite{Martinez:2012tu,Florkowski:2013lza,Bazow:2013ifa}. The fact that there must be a maximum in particle production as a function of $\eta/s$ can be anticipated by the fact that particle production should vanish in both the ideal and free streaming limits.  The behavior found using anisotropic hydrodynamics is qualitatively different than all known standard second-order viscous hydrodynamics approaches, which predict that particle production increases monotonically as $\eta/s$ increases.  The non-monotonicity of particle production becomes particularly important when enforcing fixed final multiplicity of particles, since this is typically accomplished by rescaling the initial central temperature while holding other parameters fixed.   Such a temperature rescaling can affect dilepton yields, since there is strong sensitivity of the dilepton spectra to the temperature of the emitting source.

In order to extract the freeze-out hypersurface, we parameterize space-time in the following way
\begin{eqnarray}
  t &=& \left(\tau_0 + d(\zeta,\phi,\theta) \sin\theta \sin\zeta \right)
         \cosh (d(\zeta,\phi,\theta) \cos\theta) \, , \nonumber \\
  x &=& d(\zeta,\phi,\theta) \sin\theta \cos\zeta \cos\phi \, , \nonumber \\
  y &=& d(\zeta,\phi,\theta) \sin\theta \cos\zeta \sin\phi \, , \nonumber \\
  z &=& \left(\tau_0 + d(\zeta,\phi,\theta) \sin\theta \sin\zeta \right)
          \sinh (d(\zeta,\phi,\theta) \cos\theta) \, . \hspace{6mm} 
\label{3d-par1}
\end{eqnarray}
This parametrization leads to simple formulas for the space-time rapidity $\varsigma$, longitudinal proper time $\tau$, and the
transverse distance $r$,
\begin{eqnarray}
\varsigma &=&  d(\zeta,\phi,\theta) \cos\theta  \, ,
\nonumber \\
 \tau &=& \tau_0 + d(\zeta,\phi,\theta) \sin\theta \sin\zeta  \, ,
 \nonumber \\
 r &=&  d(\zeta,\phi,\theta) \cos\zeta  \, .
\label{3d-par2}
\end{eqnarray}
The three angles $\zeta$, $\phi$, and $\theta$ are restricted to the ranges
\begin{eqnarray}
0 \leq &\,\zeta\,& \leq \pi/2 \, , \nonumber \\
0 \leq &\,\phi\,& < 2 \pi \, , \nonumber \\
0 \leq &\,\theta\,& \leq \pi \, .
\label{3d-angles}
\end{eqnarray}
The quantity $d(\zeta,\phi,\theta)$ describes the distance between a point on the freeze-out hypersurface and the coordinate
system's origin $(\tau=\tau_0,x=0,y=0,\varsigma=0)$. The parametrization (\ref{3d-par1}) works quite well for all smooth initial conditions where the distance $d$ is a function of $\zeta$, $\phi$, and $\theta$. Using the parametrization (\ref{3d-par1}), one can integrate the particle number on the freeze-out hypersurface specified by constant effective temperature $T_{\rm FO}$ in the following way
\begin{equation}
N= \int d\Sigma_\mu u^\mu n\left(T_{\rm FO} {\cal R}^{-1/4}(\xi(\zeta,\phi,\theta)),\xi(\zeta,\phi,\theta)\right)  \,  ,
\label{N}
\end{equation}
where the form of $d\Sigma_\mu$ may be obtained with the help of the formula known from differential geometry
\begin{equation}
d\Sigma_\mu = \varepsilon_{\mu \alpha \beta \gamma}
\frac{\partial x^\alpha}{\partial \zeta} \frac{\partial x^\beta}{\partial \phi} \frac{\partial x^\gamma}{\partial \theta }
d \zeta d \phi d  \theta  \, .
\label{d3Sigma}
\end{equation}
The tensor  $\varepsilon_{\mu \alpha \beta \gamma}$ is the four-index antisymmetric Levi-Civita tensor with $\varepsilon_{0123} = 1$. The quantity $d\Sigma_\mu$ defines a four-vector
that is perpendicular to the hypersurface at point $x^\mu$. Its norm is equal to the volume of the hypersurface element.
The variables $\zeta$, $\phi$, and $\theta$ introduce a coordinate system in Minkowski space parameterizing the positions of points
on the freeze-out hypersurface. Their ordering is chosen in such a way that $d\Sigma_\mu$ points in the direction of decreasing
temperature.

\begin{figure}[t]
\vspace{5mm}
\includegraphics[scale=0.71]{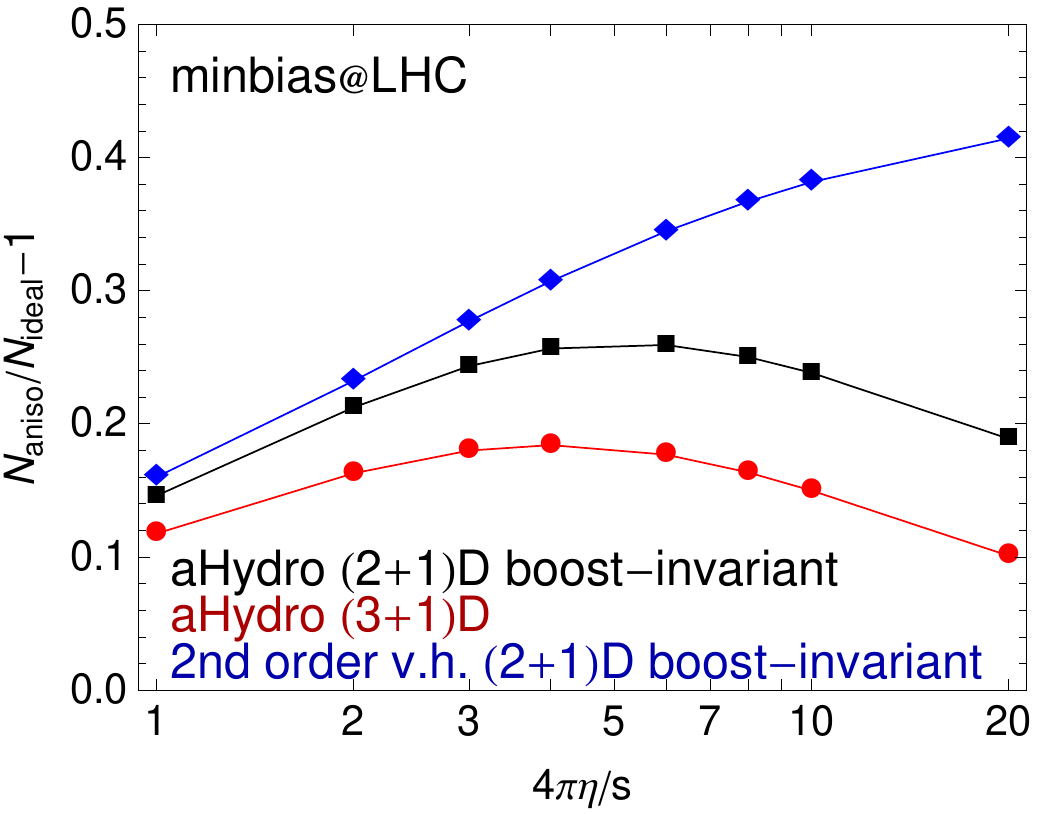}
\caption{
(Color online) The $\eta/s$ dependance of the particle production measure $N_{\rm aniso} /N_{\rm ideal} - 1$. Black squares and red dots denote the (2+1)D boost-invariant and (3+1)D anisotropic hydrodynamics, respectively. We compare them with the results obtained within (2+1)D boost-invariant viscous hydrodynamics (blue diamonds).
}
\label{fig:partprod}
\end{figure}

In Fig.~\ref{fig:partprod}, we plot the particle production measure, $N_{\rm aniso}/N_{\rm ideal} - 1$ as a function of $\bar \eta$, where $N_{\rm aniso\, (ideal)}$ denotes the density of gluons (\ref{densaniso}) integrated on the isothermal hypersurface, i.e. surface satisfying $T_{\rm eff} = T_{\rm FO} = 150$ MeV.  In Fig.~\ref{fig:partprod}, blue diamonds and black squares present the calculation for boost-invariant versions of viscous and anisotropic hydrodynamic models, respectively.  The (2+1)-dimensional viscous hydrodynamics results were generated using the code of Luzum and Romatschke \cite{Luzum:2009sb}.  We also show results for full (3+1)-dimensional anisotropic code (red dots).  We note that there is some quantitative uncertainty in the presented results due to the effective-temperature freeze-out prescription used. Another possibility for the freeze-out condition would be to use a constant value of the Knudsen number \cite{Niemi:2014wta}.  We have not considered this possibility in this work.

\bibliography{dilepton}

\begin{thebibliography}{64}%
\makeatletter
\providecommand \@ifxundefined [1]{%
 \@ifx{#1\undefined}
}%
\providecommand \@ifnum [1]{%
 \ifnum #1\expandafter \@firstoftwo
 \else \expandafter \@secondoftwo
 \fi
}%
\providecommand \@ifx [1]{%
 \ifx #1\expandafter \@firstoftwo
 \else \expandafter \@secondoftwo
 \fi
}%
\providecommand \natexlab [1]{#1}%
\providecommand \enquote  [1]{``#1''}%
\providecommand \bibnamefont  [1]{#1}%
\providecommand \bibfnamefont [1]{#1}%
\providecommand \citenamefont [1]{#1}%
\providecommand \href@noop [0]{\@secondoftwo}%
\providecommand \href [0]{\begingroup \@sanitize@url \@href}%
\providecommand \@href[1]{\@@startlink{#1}\@@href}%
\providecommand \@@href[1]{\endgroup#1\@@endlink}%
\providecommand \@sanitize@url [0]{\catcode `\\12\catcode `\$12\catcode
  `\&12\catcode `\#12\catcode `\^12\catcode `\_12\catcode `\%12\relax}%
\providecommand \@@startlink[1]{}%
\providecommand \@@endlink[0]{}%
\providecommand \url  [0]{\begingroup\@sanitize@url \@url }%
\providecommand \@url [1]{\endgroup\@href {#1}{\urlprefix }}%
\providecommand \urlprefix  [0]{URL }%
\providecommand \Eprint [0]{\href }%
\providecommand \doibase [0]{http://dx.doi.org/}%
\providecommand \selectlanguage [0]{\@gobble}%
\providecommand \bibinfo  [0]{\@secondoftwo}%
\providecommand \bibfield  [0]{\@secondoftwo}%
\providecommand \translation [1]{[#1]}%
\providecommand \BibitemOpen [0]{}%
\providecommand \bibitemStop [0]{}%
\providecommand \bibitemNoStop [0]{.\EOS\space}%
\providecommand \EOS [0]{\spacefactor3000\relax}%
\providecommand \BibitemShut  [1]{\csname bibitem#1\endcsname}%
\let\auto@bib@innerbib\@empty
\bibitem [{\citenamefont {Strickland}(2013)}]{Strickland:2013uga}%
  \BibitemOpen
  \bibfield  {author} {\bibinfo {author} {\bibfnamefont {M.}~\bibnamefont
  {Strickland}},\ }\href@noop {} {\  (\bibinfo {year} {2013})},\ \Eprint
  {http://arxiv.org/abs/1312.2285} {arXiv:1312.2285 [hep-ph]} \BibitemShut
  {NoStop}%
\bibitem [{\citenamefont {Shuryak}(1978)}]{Shuryak:1978ij}%
  \BibitemOpen
  \bibfield  {author} {\bibinfo {author} {\bibfnamefont {E.~V.}\ \bibnamefont
  {Shuryak}},\ }\href {\doibase 10.1016/0370-2693(78)90370-2} {\bibfield
  {journal} {\bibinfo  {journal} {Phys.Lett.}\ }\textbf {\bibinfo {volume}
  {B78}},\ \bibinfo {pages} {150} (\bibinfo {year} {1978})}\BibitemShut
  {NoStop}%
\bibitem [{\citenamefont {Domokos}\ and\ \citenamefont
  {Goldman}(1981)}]{Domokos:1980ba}%
  \BibitemOpen
  \bibfield  {author} {\bibinfo {author} {\bibfnamefont {G.}~\bibnamefont
  {Domokos}}\ and\ \bibinfo {author} {\bibfnamefont {J.~I.}\ \bibnamefont
  {Goldman}},\ }\href {\doibase 10.1103/PhysRevD.23.203} {\bibfield  {journal}
  {\bibinfo  {journal} {Phys.Rev.}\ }\textbf {\bibinfo {volume} {D23}},\
  \bibinfo {pages} {203} (\bibinfo {year} {1981})}\BibitemShut {NoStop}%
\bibitem [{\citenamefont {Kajantie}\ and\ \citenamefont
  {Miettinen}(1981)}]{Kajantie:1981wg}%
  \BibitemOpen
  \bibfield  {author} {\bibinfo {author} {\bibfnamefont {K.}~\bibnamefont
  {Kajantie}}\ and\ \bibinfo {author} {\bibfnamefont {H.}~\bibnamefont
  {Miettinen}},\ }\href {\doibase 10.1007/BF01548770} {\bibfield  {journal}
  {\bibinfo  {journal} {Z.Phys.}\ }\textbf {\bibinfo {volume} {C9}},\ \bibinfo
  {pages} {341} (\bibinfo {year} {1981})}\BibitemShut {NoStop}%
\bibitem [{\citenamefont {Kajantie}\ \emph {et~al.}(1986)\citenamefont
  {Kajantie}, \citenamefont {Kapusta}, \citenamefont {McLerran},\ and\
  \citenamefont {Mekjian}}]{Kajantie:1986dh}%
  \BibitemOpen
  \bibfield  {author} {\bibinfo {author} {\bibfnamefont {K.}~\bibnamefont
  {Kajantie}}, \bibinfo {author} {\bibfnamefont {J.~I.}\ \bibnamefont
  {Kapusta}}, \bibinfo {author} {\bibfnamefont {L.~D.}\ \bibnamefont
  {McLerran}}, \ and\ \bibinfo {author} {\bibfnamefont {A.}~\bibnamefont
  {Mekjian}},\ }\href {\doibase 10.1103/PhysRevD.34.2746} {\bibfield  {journal}
  {\bibinfo  {journal} {Phys.Rev.}\ }\textbf {\bibinfo {volume} {D34}},\
  \bibinfo {pages} {2746} (\bibinfo {year} {1986})}\BibitemShut {NoStop}%
\bibitem [{\citenamefont {Kapusta}\ \emph {et~al.}(1992)\citenamefont
  {Kapusta}, \citenamefont {McLerran},\ and\ \citenamefont
  {Kumar~Srivastava}}]{Kapusta:1992uy}%
  \BibitemOpen
  \bibfield  {author} {\bibinfo {author} {\bibfnamefont {J.~I.}\ \bibnamefont
  {Kapusta}}, \bibinfo {author} {\bibfnamefont {L.~D.}\ \bibnamefont
  {McLerran}}, \ and\ \bibinfo {author} {\bibfnamefont {D.}~\bibnamefont
  {Kumar~Srivastava}},\ }\href {\doibase 10.1016/0370-2693(92)91445-F}
  {\bibfield  {journal} {\bibinfo  {journal} {Phys. Lett.}\ }\textbf {\bibinfo
  {volume} {B283}},\ \bibinfo {pages} {145} (\bibinfo {year}
  {1992})}\BibitemShut {NoStop}%
\bibitem [{\citenamefont {Strickland}(1994)}]{Strickland:1994rf}%
  \BibitemOpen
  \bibfield  {author} {\bibinfo {author} {\bibfnamefont {M.}~\bibnamefont
  {Strickland}},\ }\href {\doibase 10.1016/0370-2693(94)91045-6} {\bibfield
  {journal} {\bibinfo  {journal} {Phys.Lett.}\ }\textbf {\bibinfo {volume}
  {B331}},\ \bibinfo {pages} {245} (\bibinfo {year} {1994})}\BibitemShut
  {NoStop}%
\bibitem [{\citenamefont {Rapp}\ and\ \citenamefont
  {Wambach}(2000)}]{Rapp:1999ej}%
  \BibitemOpen
  \bibfield  {author} {\bibinfo {author} {\bibfnamefont {R.}~\bibnamefont
  {Rapp}}\ and\ \bibinfo {author} {\bibfnamefont {J.}~\bibnamefont {Wambach}},\
  }\href {\doibase 10.1007/0-306-47101-9_1} {\bibfield  {journal} {\bibinfo
  {journal} {Adv.Nucl.Phys.}\ }\textbf {\bibinfo {volume} {25}},\ \bibinfo
  {pages} {1} (\bibinfo {year} {2000})},\ \Eprint
  {http://arxiv.org/abs/hep-ph/9909229} {arXiv:hep-ph/9909229 [hep-ph]}
  \BibitemShut {NoStop}%
\bibitem [{\citenamefont {Rapp}\ and\ \citenamefont
  {Shuryak}(2000)}]{Rapp:1999zw}%
  \BibitemOpen
  \bibfield  {author} {\bibinfo {author} {\bibfnamefont {R.}~\bibnamefont
  {Rapp}}\ and\ \bibinfo {author} {\bibfnamefont {E.~V.}\ \bibnamefont
  {Shuryak}},\ }\href {\doibase 10.1016/S0370-2693(99)01367-2} {\bibfield
  {journal} {\bibinfo  {journal} {Phys.Lett.}\ }\textbf {\bibinfo {volume}
  {B473}},\ \bibinfo {pages} {13} (\bibinfo {year} {2000})},\ \Eprint
  {http://arxiv.org/abs/hep-ph/9909348} {arXiv:hep-ph/9909348 [hep-ph]}
  \BibitemShut {NoStop}%
\bibitem [{\citenamefont {Rapp}(2001)}]{Rapp:2000pe}%
  \BibitemOpen
  \bibfield  {author} {\bibinfo {author} {\bibfnamefont {R.}~\bibnamefont
  {Rapp}},\ }\href {\doibase 10.1103/PhysRevC.63.054907,
  10.1103/PhysRevA.83.043833} {\bibfield  {journal} {\bibinfo  {journal}
  {Phys.Rev.}\ }\textbf {\bibinfo {volume} {C63}},\ \bibinfo {pages} {054907}
  (\bibinfo {year} {2001})},\ \Eprint {http://arxiv.org/abs/hep-ph/0010101}
  {arXiv:hep-ph/0010101 [hep-ph]} \BibitemShut {NoStop}%
\bibitem [{\citenamefont {Arnold}\ \emph
  {et~al.}(2001{\natexlab{a}})\citenamefont {Arnold}, \citenamefont {Moore},\
  and\ \citenamefont {Yaffe}}]{Arnold:2001ba}%
  \BibitemOpen
  \bibfield  {author} {\bibinfo {author} {\bibfnamefont {P.}~\bibnamefont
  {Arnold}}, \bibinfo {author} {\bibfnamefont {G.~D.}\ \bibnamefont {Moore}}, \
  and\ \bibinfo {author} {\bibfnamefont {L.~G.}\ \bibnamefont {Yaffe}},\
  }\href@noop {} {\bibfield  {journal} {\bibinfo  {journal} {JHEP}\ }\textbf
  {\bibinfo {volume} {11}},\ \bibinfo {pages} {057} (\bibinfo {year}
  {2001}{\natexlab{a}})},\ \Eprint {http://arxiv.org/abs/hep-ph/0109064}
  {hep-ph/0109064} \BibitemShut {NoStop}%
\bibitem [{\citenamefont {Arnold}\ \emph
  {et~al.}(2001{\natexlab{b}})\citenamefont {Arnold}, \citenamefont {Moore},\
  and\ \citenamefont {Yaffe}}]{Arnold:2001ms}%
  \BibitemOpen
  \bibfield  {author} {\bibinfo {author} {\bibfnamefont {P.}~\bibnamefont
  {Arnold}}, \bibinfo {author} {\bibfnamefont {G.~D.}\ \bibnamefont {Moore}}, \
  and\ \bibinfo {author} {\bibfnamefont {L.~G.}\ \bibnamefont {Yaffe}},\
  }\href@noop {} {\bibfield  {journal} {\bibinfo  {journal} {JHEP}\ }\textbf
  {\bibinfo {volume} {12}},\ \bibinfo {pages} {009} (\bibinfo {year}
  {2001}{\natexlab{b}})},\ \Eprint {http://arxiv.org/abs/hep-ph/0111107}
  {hep-ph/0111107} \BibitemShut {NoStop}%
\bibitem [{\citenamefont {Aurenche}\ \emph {et~al.}(2002)\citenamefont
  {Aurenche}, \citenamefont {Gelis}, \citenamefont {Moore},\ and\ \citenamefont
  {Zaraket}}]{Aurenche:2002wq}%
  \BibitemOpen
  \bibfield  {author} {\bibinfo {author} {\bibfnamefont {P.}~\bibnamefont
  {Aurenche}}, \bibinfo {author} {\bibfnamefont {F.}~\bibnamefont {Gelis}},
  \bibinfo {author} {\bibfnamefont {G.}~\bibnamefont {Moore}}, \ and\ \bibinfo
  {author} {\bibfnamefont {H.}~\bibnamefont {Zaraket}},\ }\href {\doibase
  10.1088/1126-6708/2002/12/006} {\bibfield  {journal} {\bibinfo  {journal}
  {JHEP}\ }\textbf {\bibinfo {volume} {0212}},\ \bibinfo {pages} {006}
  (\bibinfo {year} {2002})},\ \Eprint {http://arxiv.org/abs/hep-ph/0211036}
  {arXiv:hep-ph/0211036 [hep-ph]} \BibitemShut {NoStop}%
\bibitem [{\citenamefont {Arnold}\ \emph {et~al.}(2002)\citenamefont {Arnold},
  \citenamefont {Moore},\ and\ \citenamefont {Yaffe}}]{Arnold:2002ja}%
  \BibitemOpen
  \bibfield  {author} {\bibinfo {author} {\bibfnamefont {P.~B.}\ \bibnamefont
  {Arnold}}, \bibinfo {author} {\bibfnamefont {G.~D.}\ \bibnamefont {Moore}}, \
  and\ \bibinfo {author} {\bibfnamefont {L.~G.}\ \bibnamefont {Yaffe}},\ }\href
  {\doibase 10.1088/1126-6708/2002/06/030} {\bibfield  {journal} {\bibinfo
  {journal} {JHEP}\ }\textbf {\bibinfo {volume} {0206}},\ \bibinfo {pages}
  {030} (\bibinfo {year} {2002})},\ \Eprint
  {http://arxiv.org/abs/hep-ph/0204343} {arXiv:hep-ph/0204343 [hep-ph]}
  \BibitemShut {NoStop}%
\bibitem [{\citenamefont {Dusling}\ and\ \citenamefont
  {Lin}(2008)}]{Dusling:2008xj}%
  \BibitemOpen
  \bibfield  {author} {\bibinfo {author} {\bibfnamefont {K.}~\bibnamefont
  {Dusling}}\ and\ \bibinfo {author} {\bibfnamefont {S.}~\bibnamefont {Lin}},\
  }\href {\doibase 10.1016/j.nuclphysa.2008.06.007} {\bibfield  {journal}
  {\bibinfo  {journal} {Nucl.Phys.}\ }\textbf {\bibinfo {volume} {A809}},\
  \bibinfo {pages} {246} (\bibinfo {year} {2008})},\ \Eprint
  {http://arxiv.org/abs/0803.1262} {arXiv:0803.1262 [nucl-th]} \BibitemShut
  {NoStop}%
\bibitem [{\citenamefont {Vujanovic}\ \emph {et~al.}(2014)\citenamefont
  {Vujanovic}, \citenamefont {Young}, \citenamefont {Schenke}, \citenamefont
  {Rapp}, \citenamefont {Jeon} \emph {et~al.}}]{Vujanovic:2013jpa}%
  \BibitemOpen
  \bibfield  {author} {\bibinfo {author} {\bibfnamefont {G.}~\bibnamefont
  {Vujanovic}}, \bibinfo {author} {\bibfnamefont {C.}~\bibnamefont {Young}},
  \bibinfo {author} {\bibfnamefont {B.}~\bibnamefont {Schenke}}, \bibinfo
  {author} {\bibfnamefont {R.}~\bibnamefont {Rapp}}, \bibinfo {author}
  {\bibfnamefont {S.}~\bibnamefont {Jeon}},  \emph {et~al.},\ }\href {\doibase
  10.1103/PhysRevC.89.034904} {\bibfield  {journal} {\bibinfo  {journal}
  {Phys.Rev.}\ }\textbf {\bibinfo {volume} {C89}},\ \bibinfo {pages} {034904}
  (\bibinfo {year} {2014})},\ \Eprint {http://arxiv.org/abs/1312.0676}
  {arXiv:1312.0676 [nucl-th]} \BibitemShut {NoStop}%
\bibitem [{\citenamefont {Endres}\ \emph {et~al.}(2014)\citenamefont {Endres},
  \citenamefont {van Hees}, \citenamefont {Weil},\ and\ \citenamefont
  {Bleicher}}]{Endres:2014zua}%
  \BibitemOpen
  \bibfield  {author} {\bibinfo {author} {\bibfnamefont {S.}~\bibnamefont
  {Endres}}, \bibinfo {author} {\bibfnamefont {H.}~\bibnamefont {van Hees}},
  \bibinfo {author} {\bibfnamefont {J.}~\bibnamefont {Weil}}, \ and\ \bibinfo
  {author} {\bibfnamefont {M.}~\bibnamefont {Bleicher}},\ }\href@noop {} {\
  (\bibinfo {year} {2014})},\ \Eprint {http://arxiv.org/abs/1412.1965}
  {arXiv:1412.1965 [nucl-th]} \BibitemShut {NoStop}%
\bibitem [{\citenamefont {Rapp}(2013)}]{Rapp:2013nxa}%
  \BibitemOpen
  \bibfield  {author} {\bibinfo {author} {\bibfnamefont {R.}~\bibnamefont
  {Rapp}},\ }\href {\doibase 10.1155/2013/148253} {\bibfield  {journal}
  {\bibinfo  {journal} {Adv.High Energy Phys.}\ }\textbf {\bibinfo {volume}
  {2013}},\ \bibinfo {pages} {148253} (\bibinfo {year} {2013})},\ \Eprint
  {http://arxiv.org/abs/1304.2309} {arXiv:1304.2309 [hep-ph]} \BibitemShut
  {NoStop}%
\bibitem [{\citenamefont {Sakaguchi}(2014)}]{Sakaguchi:2014ewa}%
  \BibitemOpen
  \bibfield  {author} {\bibinfo {author} {\bibfnamefont {T.}~\bibnamefont
  {Sakaguchi}},\ }\href@noop {} {\  (\bibinfo {year} {2014})},\ \Eprint
  {http://arxiv.org/abs/1401.2481} {arXiv:1401.2481 [nucl-ex]} \BibitemShut
  {NoStop}%
\bibitem [{\citenamefont {Martinez}\ and\ \citenamefont
  {Strickland}(2008{\natexlab{a}})}]{Mauricio:2007vz}%
  \BibitemOpen
  \bibfield  {author} {\bibinfo {author} {\bibfnamefont {M.}~\bibnamefont
  {Martinez}}\ and\ \bibinfo {author} {\bibfnamefont {M.}~\bibnamefont
  {Strickland}},\ }\href@noop {} {\bibfield  {journal} {\bibinfo  {journal}
  {Phys.Rev.Lett.}\ }\textbf {\bibinfo {volume} {100}},\ \bibinfo {pages}
  {102301} (\bibinfo {year} {2008}{\natexlab{a}})},\ \Eprint
  {http://arxiv.org/abs/0709.3576} {arXiv:0709.3576 [hep-ph]} \BibitemShut
  {NoStop}%
\bibitem [{\citenamefont {Martinez}\ and\ \citenamefont
  {Strickland}(2008{\natexlab{b}})}]{Martinez:2008di}%
  \BibitemOpen
  \bibfield  {author} {\bibinfo {author} {\bibfnamefont {M.}~\bibnamefont
  {Martinez}}\ and\ \bibinfo {author} {\bibfnamefont {M.}~\bibnamefont
  {Strickland}},\ }\href {\doibase 10.1103/PhysRevC.78.034917} {\bibfield
  {journal} {\bibinfo  {journal} {Phys.Rev.}\ }\textbf {\bibinfo {volume}
  {C78}},\ \bibinfo {pages} {034917} (\bibinfo {year} {2008}{\natexlab{b}})},\
  \Eprint {http://arxiv.org/abs/0805.4552} {arXiv:0805.4552 [hep-ph]}
  \BibitemShut {NoStop}%
\bibitem [{\citenamefont {Martinez}\ and\ \citenamefont
  {Strickland}(2010)}]{Martinez:2010sc}%
  \BibitemOpen
  \bibfield  {author} {\bibinfo {author} {\bibfnamefont {M.}~\bibnamefont
  {Martinez}}\ and\ \bibinfo {author} {\bibfnamefont {M.}~\bibnamefont
  {Strickland}},\ }\href {\doibase 10.1016/j.nuclphysa.2010.08.011} {\bibfield
  {journal} {\bibinfo  {journal} {Nucl. Phys.}\ }\textbf {\bibinfo {volume}
  {A848}},\ \bibinfo {pages} {183} (\bibinfo {year} {2010})},\ \Eprint
  {http://arxiv.org/abs/1007.0889} {arXiv:1007.0889 [nucl-th]} \BibitemShut
  {NoStop}%
\bibitem [{\citenamefont {Florkowski}\ and\ \citenamefont
  {Ryblewski}(2011)}]{Florkowski:2010cf}%
  \BibitemOpen
  \bibfield  {author} {\bibinfo {author} {\bibfnamefont {W.}~\bibnamefont
  {Florkowski}}\ and\ \bibinfo {author} {\bibfnamefont {R.}~\bibnamefont
  {Ryblewski}},\ }\href {\doibase 10.1103/PhysRevC.83.034907} {\bibfield
  {journal} {\bibinfo  {journal} {Phys.Rev.}\ }\textbf {\bibinfo {volume}
  {C83}},\ \bibinfo {pages} {034907} (\bibinfo {year} {2011})},\ \Eprint
  {http://arxiv.org/abs/1007.0130} {arXiv:1007.0130 [nucl-th]} \BibitemShut
  {NoStop}%
\bibitem [{\citenamefont {Ryblewski}\ and\ \citenamefont
  {Florkowski}(2011{\natexlab{a}})}]{Ryblewski:2010bs}%
  \BibitemOpen
  \bibfield  {author} {\bibinfo {author} {\bibfnamefont {R.}~\bibnamefont
  {Ryblewski}}\ and\ \bibinfo {author} {\bibfnamefont {W.}~\bibnamefont
  {Florkowski}},\ }\href {\doibase 10.1088/0954-3899/38/1/015104} {\bibfield
  {journal} {\bibinfo  {journal} {J.Phys.G}\ }\textbf {\bibinfo {volume}
  {G38}},\ \bibinfo {pages} {015104} (\bibinfo {year} {2011}{\natexlab{a}})},\
  \Eprint {http://arxiv.org/abs/1007.4662} {arXiv:1007.4662 [nucl-th]}
  \BibitemShut {NoStop}%
\bibitem [{\citenamefont {Martinez}\ and\ \citenamefont
  {Strickland}(2011)}]{Martinez:2010sd}%
  \BibitemOpen
  \bibfield  {author} {\bibinfo {author} {\bibfnamefont {M.}~\bibnamefont
  {Martinez}}\ and\ \bibinfo {author} {\bibfnamefont {M.}~\bibnamefont
  {Strickland}},\ }\href {\doibase 10.1016/j.nuclphysa.2011.02.003} {\bibfield
  {journal} {\bibinfo  {journal} {Nucl.Phys.}\ }\textbf {\bibinfo {volume}
  {A856}},\ \bibinfo {pages} {68} (\bibinfo {year} {2011})},\ \Eprint
  {http://arxiv.org/abs/1011.3056} {arXiv:1011.3056 [nucl-th]} \BibitemShut
  {NoStop}%
\bibitem [{\citenamefont {Ryblewski}\ and\ \citenamefont
  {Florkowski}(2011{\natexlab{b}})}]{Ryblewski:2011aq}%
  \BibitemOpen
  \bibfield  {author} {\bibinfo {author} {\bibfnamefont {R.}~\bibnamefont
  {Ryblewski}}\ and\ \bibinfo {author} {\bibfnamefont {W.}~\bibnamefont
  {Florkowski}},\ }\href {\doibase 10.1140/epjc/s10052-011-1761-8} {\bibfield
  {journal} {\bibinfo  {journal} {Eur.Phys.J.}\ }\textbf {\bibinfo {volume}
  {C71}},\ \bibinfo {pages} {1761} (\bibinfo {year} {2011}{\natexlab{b}})},\
  \Eprint {http://arxiv.org/abs/1103.1260} {arXiv:1103.1260 [nucl-th]}
  \BibitemShut {NoStop}%
\bibitem [{\citenamefont {Florkowski}\ and\ \citenamefont
  {Ryblewski}(2012)}]{Florkowski:2011jg}%
  \BibitemOpen
  \bibfield  {author} {\bibinfo {author} {\bibfnamefont {W.}~\bibnamefont
  {Florkowski}}\ and\ \bibinfo {author} {\bibfnamefont {R.}~\bibnamefont
  {Ryblewski}},\ }\href {\doibase 10.1103/PhysRevC.85.044902} {\bibfield
  {journal} {\bibinfo  {journal} {Phys.Rev.}\ }\textbf {\bibinfo {volume}
  {C85}},\ \bibinfo {pages} {044902} (\bibinfo {year} {2012})},\ \Eprint
  {http://arxiv.org/abs/1111.5997} {arXiv:1111.5997 [nucl-th]} \BibitemShut
  {NoStop}%
\bibitem [{\citenamefont {Martinez}\ \emph {et~al.}(2012)\citenamefont
  {Martinez}, \citenamefont {Ryblewski},\ and\ \citenamefont
  {Strickland}}]{Martinez:2012tu}%
  \BibitemOpen
  \bibfield  {author} {\bibinfo {author} {\bibfnamefont {M.}~\bibnamefont
  {Martinez}}, \bibinfo {author} {\bibfnamefont {R.}~\bibnamefont {Ryblewski}},
  \ and\ \bibinfo {author} {\bibfnamefont {M.}~\bibnamefont {Strickland}},\
  }\href@noop {} {\bibfield  {journal} {\bibinfo  {journal} {Phys.Rev.}\
  }\textbf {\bibinfo {volume} {C85}},\ \bibinfo {pages} {064913} (\bibinfo
  {year} {2012})},\ \Eprint {http://arxiv.org/abs/1204.1473} {arXiv:1204.1473
  [nucl-th]} \BibitemShut {NoStop}%
\bibitem [{\citenamefont {Ryblewski}\ and\ \citenamefont
  {Florkowski}(2012)}]{Ryblewski:2012rr}%
  \BibitemOpen
  \bibfield  {author} {\bibinfo {author} {\bibfnamefont {R.}~\bibnamefont
  {Ryblewski}}\ and\ \bibinfo {author} {\bibfnamefont {W.}~\bibnamefont
  {Florkowski}},\ }\href@noop {} {\bibfield  {journal} {\bibinfo  {journal}
  {Phys.Rev.}\ }\textbf {\bibinfo {volume} {C85}},\ \bibinfo {pages} {064901}
  (\bibinfo {year} {2012})},\ \Eprint {http://arxiv.org/abs/1204.2624}
  {arXiv:1204.2624 [nucl-th]} \BibitemShut {NoStop}%
\bibitem [{\citenamefont {Florkowski}\ \emph
  {et~al.}(2013{\natexlab{a}})\citenamefont {Florkowski}, \citenamefont {Maj},
  \citenamefont {Ryblewski},\ and\ \citenamefont
  {Strickland}}]{Florkowski:2012as}%
  \BibitemOpen
  \bibfield  {author} {\bibinfo {author} {\bibfnamefont {W.}~\bibnamefont
  {Florkowski}}, \bibinfo {author} {\bibfnamefont {R.}~\bibnamefont {Maj}},
  \bibinfo {author} {\bibfnamefont {R.}~\bibnamefont {Ryblewski}}, \ and\
  \bibinfo {author} {\bibfnamefont {M.}~\bibnamefont {Strickland}},\ }\href
  {\doibase 10.1103/PhysRevC.87.034914} {\bibfield  {journal} {\bibinfo
  {journal} {Phys.Rev.}\ }\textbf {\bibinfo {volume} {C87}},\ \bibinfo {pages}
  {034914} (\bibinfo {year} {2013}{\natexlab{a}})},\ \Eprint
  {http://arxiv.org/abs/1209.3671} {arXiv:1209.3671 [nucl-th]} \BibitemShut
  {NoStop}%
\bibitem [{\citenamefont {Florkowski}\ and\ \citenamefont
  {Maj}(2013)}]{Florkowski:2013uqa}%
  \BibitemOpen
  \bibfield  {author} {\bibinfo {author} {\bibfnamefont {W.}~\bibnamefont
  {Florkowski}}\ and\ \bibinfo {author} {\bibfnamefont {R.}~\bibnamefont
  {Maj}},\ }\href {\doibase 10.5506/APhysPolB.44.2003} {\bibfield  {journal}
  {\bibinfo  {journal} {Acta Phys.Polon.}\ }\textbf {\bibinfo {volume} {B44}},\
  \bibinfo {pages} {2003} (\bibinfo {year} {2013})},\ \Eprint
  {http://arxiv.org/abs/1309.2786} {arXiv:1309.2786 [nucl-th]} \BibitemShut
  {NoStop}%
\bibitem [{\citenamefont {Ryblewski}(2013)}]{Ryblewski:2013jsa}%
  \BibitemOpen
  \bibfield  {author} {\bibinfo {author} {\bibfnamefont {R.}~\bibnamefont
  {Ryblewski}},\ }\href {\doibase 10.1088/0954-3899/40/9/093101} {\bibfield
  {journal} {\bibinfo  {journal} {J.Phys.}\ }\textbf {\bibinfo {volume}
  {G40}},\ \bibinfo {pages} {093101} (\bibinfo {year} {2013})}\BibitemShut
  {NoStop}%
\bibitem [{\citenamefont {Florkowski}\ \emph
  {et~al.}(2013{\natexlab{b}})\citenamefont {Florkowski}, \citenamefont
  {Ryblewski},\ and\ \citenamefont {Strickland}}]{Florkowski:2013lza}%
  \BibitemOpen
  \bibfield  {author} {\bibinfo {author} {\bibfnamefont {W.}~\bibnamefont
  {Florkowski}}, \bibinfo {author} {\bibfnamefont {R.}~\bibnamefont
  {Ryblewski}}, \ and\ \bibinfo {author} {\bibfnamefont {M.}~\bibnamefont
  {Strickland}},\ }\href {\doibase 10.1016/j.nuclphysa.2013.08.004} {\bibfield
  {journal} {\bibinfo  {journal} {Nucl.Phys.}\ }\textbf {\bibinfo {volume}
  {A916}},\ \bibinfo {pages} {249} (\bibinfo {year} {2013}{\natexlab{b}})},\
  \Eprint {http://arxiv.org/abs/1304.0665} {arXiv:1304.0665 [nucl-th]}
  \BibitemShut {NoStop}%
\bibitem [{\citenamefont {Bazow}\ \emph {et~al.}(2014)\citenamefont {Bazow},
  \citenamefont {Heinz},\ and\ \citenamefont {Strickland}}]{Bazow:2013ifa}%
  \BibitemOpen
  \bibfield  {author} {\bibinfo {author} {\bibfnamefont {D.}~\bibnamefont
  {Bazow}}, \bibinfo {author} {\bibfnamefont {U.~W.}\ \bibnamefont {Heinz}}, \
  and\ \bibinfo {author} {\bibfnamefont {M.}~\bibnamefont {Strickland}},\
  }\href {\doibase 10.1103/PhysRevC.90.054910} {\bibfield  {journal} {\bibinfo
  {journal} {Phys.Rev.}\ }\textbf {\bibinfo {volume} {C90}},\ \bibinfo {pages}
  {044908} (\bibinfo {year} {2014})},\ \Eprint {http://arxiv.org/abs/1311.6720}
  {arXiv:1311.6720 [nucl-th]} \BibitemShut {NoStop}%
\bibitem [{\citenamefont {Tinti}\ and\ \citenamefont
  {Florkowski}(2014)}]{Tinti:2013vba}%
  \BibitemOpen
  \bibfield  {author} {\bibinfo {author} {\bibfnamefont {L.}~\bibnamefont
  {Tinti}}\ and\ \bibinfo {author} {\bibfnamefont {W.}~\bibnamefont
  {Florkowski}},\ }\href {\doibase 10.1103/PhysRevC.89.034907} {\bibfield
  {journal} {\bibinfo  {journal} {Phys.Rev.}\ }\textbf {\bibinfo {volume}
  {C89}},\ \bibinfo {pages} {034907} (\bibinfo {year} {2014})},\ \Eprint
  {http://arxiv.org/abs/1312.6614} {arXiv:1312.6614 [nucl-th]} \BibitemShut
  {NoStop}%
\bibitem [{\citenamefont {Florkowski}\ \emph
  {et~al.}(2014{\natexlab{a}})\citenamefont {Florkowski}, \citenamefont
  {Ryblewski}, \citenamefont {Strickland},\ and\ \citenamefont
  {Tinti}}]{Florkowski:2014bba}%
  \BibitemOpen
  \bibfield  {author} {\bibinfo {author} {\bibfnamefont {W.}~\bibnamefont
  {Florkowski}}, \bibinfo {author} {\bibfnamefont {R.}~\bibnamefont
  {Ryblewski}}, \bibinfo {author} {\bibfnamefont {M.}~\bibnamefont
  {Strickland}}, \ and\ \bibinfo {author} {\bibfnamefont {L.}~\bibnamefont
  {Tinti}},\ }\href {\doibase 10.1103/PhysRevC.89.054909} {\bibfield  {journal}
  {\bibinfo  {journal} {Phys.Rev.}\ }\textbf {\bibinfo {volume} {C89}},\
  \bibinfo {pages} {054909} (\bibinfo {year} {2014}{\natexlab{a}})},\ \Eprint
  {http://arxiv.org/abs/1403.1223} {arXiv:1403.1223 [hep-ph]} \BibitemShut
  {NoStop}%
\bibitem [{\citenamefont {Florkowski}\ and\ \citenamefont
  {Madetko}(2014)}]{Florkowski:2014txa}%
  \BibitemOpen
  \bibfield  {author} {\bibinfo {author} {\bibfnamefont {W.}~\bibnamefont
  {Florkowski}}\ and\ \bibinfo {author} {\bibfnamefont {O.}~\bibnamefont
  {Madetko}},\ }\href {\doibase 10.5506/APhysPolB.45.1103} {\bibfield
  {journal} {\bibinfo  {journal} {Acta Phys.Polon.}\ }\textbf {\bibinfo
  {volume} {B45}},\ \bibinfo {pages} {1103} (\bibinfo {year} {2014})},\ \Eprint
  {http://arxiv.org/abs/1402.2401} {arXiv:1402.2401 [nucl-th]} \BibitemShut
  {NoStop}%
\bibitem [{\citenamefont {Nopoush}\ \emph
  {et~al.}(2014{\natexlab{a}})\citenamefont {Nopoush}, \citenamefont
  {Ryblewski},\ and\ \citenamefont {Strickland}}]{Nopoush:2014pfa}%
  \BibitemOpen
  \bibfield  {author} {\bibinfo {author} {\bibfnamefont {M.}~\bibnamefont
  {Nopoush}}, \bibinfo {author} {\bibfnamefont {R.}~\bibnamefont {Ryblewski}},
  \ and\ \bibinfo {author} {\bibfnamefont {M.}~\bibnamefont {Strickland}},\
  }\href {\doibase 10.1103/PhysRevC.90.014908} {\bibfield  {journal} {\bibinfo
  {journal} {Phys.Rev.}\ }\textbf {\bibinfo {volume} {C90}},\ \bibinfo {pages}
  {014908} (\bibinfo {year} {2014}{\natexlab{a}})},\ \Eprint
  {http://arxiv.org/abs/1405.1355} {arXiv:1405.1355 [hep-ph]} \BibitemShut
  {NoStop}%
\bibitem [{\citenamefont {Denicol}\ \emph
  {et~al.}(2014{\natexlab{a}})\citenamefont {Denicol}, \citenamefont
  {Florkowski}, \citenamefont {Ryblewski},\ and\ \citenamefont
  {Strickland}}]{Denicol:2014mca}%
  \BibitemOpen
  \bibfield  {author} {\bibinfo {author} {\bibfnamefont {G.~S.}\ \bibnamefont
  {Denicol}}, \bibinfo {author} {\bibfnamefont {W.}~\bibnamefont {Florkowski}},
  \bibinfo {author} {\bibfnamefont {R.}~\bibnamefont {Ryblewski}}, \ and\
  \bibinfo {author} {\bibfnamefont {M.}~\bibnamefont {Strickland}},\ }\href
  {\doibase 10.1103/PhysRevC.90.044905} {\bibfield  {journal} {\bibinfo
  {journal} {Phys.Rev.}\ }\textbf {\bibinfo {volume} {C90}},\ \bibinfo {pages}
  {044905} (\bibinfo {year} {2014}{\natexlab{a}})},\ \Eprint
  {http://arxiv.org/abs/1407.4767} {arXiv:1407.4767 [hep-ph]} \BibitemShut
  {NoStop}%
\bibitem [{\citenamefont {Nopoush}\ \emph
  {et~al.}(2014{\natexlab{b}})\citenamefont {Nopoush}, \citenamefont
  {Ryblewski},\ and\ \citenamefont {Strickland}}]{Nopoush:2014qba}%
  \BibitemOpen
  \bibfield  {author} {\bibinfo {author} {\bibfnamefont {M.}~\bibnamefont
  {Nopoush}}, \bibinfo {author} {\bibfnamefont {R.}~\bibnamefont {Ryblewski}},
  \ and\ \bibinfo {author} {\bibfnamefont {M.}~\bibnamefont {Strickland}},\
  }\href@noop {} {\  (\bibinfo {year} {2014}{\natexlab{b}})},\ \Eprint
  {http://arxiv.org/abs/1410.6790} {arXiv:1410.6790 [nucl-th]} \BibitemShut
  {NoStop}%
\bibitem [{\citenamefont {Tinti}(2014)}]{Tinti:2014yya}%
  \BibitemOpen
  \bibfield  {author} {\bibinfo {author} {\bibfnamefont {L.}~\bibnamefont
  {Tinti}},\ }\href@noop {} {\  (\bibinfo {year} {2014})},\ \Eprint
  {http://arxiv.org/abs/1411.7268} {arXiv:1411.7268 [nucl-th]} \BibitemShut
  {NoStop}%
\bibitem [{\citenamefont {Strickland}(2014)}]{Strickland:2014pga}%
  \BibitemOpen
  \bibfield  {author} {\bibinfo {author} {\bibfnamefont {M.}~\bibnamefont
  {Strickland}},\ }\href@noop {} {\  (\bibinfo {year} {2014})},\ \Eprint
  {http://arxiv.org/abs/1410.5786} {arXiv:1410.5786 [nucl-th]} \BibitemShut
  {NoStop}%
\bibitem [{\citenamefont {Romatschke}\ and\ \citenamefont
  {Strickland}(2003)}]{Romatschke:2003ms}%
  \BibitemOpen
  \bibfield  {author} {\bibinfo {author} {\bibfnamefont {P.}~\bibnamefont
  {Romatschke}}\ and\ \bibinfo {author} {\bibfnamefont {M.}~\bibnamefont
  {Strickland}},\ }\href@noop {} {\bibfield  {journal} {\bibinfo  {journal}
  {Phys. Rev.}\ }\textbf {\bibinfo {volume} {D68}},\ \bibinfo {pages} {036004}
  (\bibinfo {year} {2003})},\ \Eprint {http://arxiv.org/abs/hep-ph/0304092}
  {hep-ph/0304092} \BibitemShut {NoStop}%
\bibitem [{\citenamefont {Romatschke}\ and\ \citenamefont
  {Strickland}(2004)}]{Romatschke:2004jh}%
  \BibitemOpen
  \bibfield  {author} {\bibinfo {author} {\bibfnamefont {P.}~\bibnamefont
  {Romatschke}}\ and\ \bibinfo {author} {\bibfnamefont {M.}~\bibnamefont
  {Strickland}},\ }\href {\doibase 10.1103/PhysRevD.70.116006} {\bibfield
  {journal} {\bibinfo  {journal} {Phys.Rev.}\ }\textbf {\bibinfo {volume}
  {D70}},\ \bibinfo {pages} {116006} (\bibinfo {year} {2004})},\ \Eprint
  {http://arxiv.org/abs/hep-ph/0406188} {arXiv:hep-ph/0406188 [hep-ph]}
  \BibitemShut {NoStop}%
\bibitem [{\citenamefont {Florkowski}\ \emph
  {et~al.}(2013{\natexlab{c}})\citenamefont {Florkowski}, \citenamefont
  {Ryblewski},\ and\ \citenamefont {Strickland}}]{Florkowski:2013lya}%
  \BibitemOpen
  \bibfield  {author} {\bibinfo {author} {\bibfnamefont {W.}~\bibnamefont
  {Florkowski}}, \bibinfo {author} {\bibfnamefont {R.}~\bibnamefont
  {Ryblewski}}, \ and\ \bibinfo {author} {\bibfnamefont {M.}~\bibnamefont
  {Strickland}},\ }\href {\doibase 10.1103/PhysRevC.88.024903} {\bibfield
  {journal} {\bibinfo  {journal} {Phys. Rev.}\ }\textbf {\bibinfo {volume}
  {C88}},\ \bibinfo {pages} {024903} (\bibinfo {year} {2013}{\natexlab{c}})},\
  \Eprint {http://arxiv.org/abs/1305.7234} {arXiv:1305.7234 [nucl-th]}
  \BibitemShut {NoStop}%
\bibitem [{\citenamefont {Florkowski}\ \emph
  {et~al.}(2014{\natexlab{b}})\citenamefont {Florkowski}, \citenamefont
  {Maksymiuk}, \citenamefont {Ryblewski},\ and\ \citenamefont
  {Strickland}}]{Florkowski:2014sfa}%
  \BibitemOpen
  \bibfield  {author} {\bibinfo {author} {\bibfnamefont {W.}~\bibnamefont
  {Florkowski}}, \bibinfo {author} {\bibfnamefont {E.}~\bibnamefont
  {Maksymiuk}}, \bibinfo {author} {\bibfnamefont {R.}~\bibnamefont
  {Ryblewski}}, \ and\ \bibinfo {author} {\bibfnamefont {M.}~\bibnamefont
  {Strickland}},\ }\href {\doibase 10.1103/PhysRevC.89.054908} {\bibfield
  {journal} {\bibinfo  {journal} {Phys.Rev.}\ }\textbf {\bibinfo {volume}
  {C89}},\ \bibinfo {pages} {054908} (\bibinfo {year} {2014}{\natexlab{b}})},\
  \Eprint {http://arxiv.org/abs/1402.7348} {arXiv:1402.7348 [hep-ph]}
  \BibitemShut {NoStop}%
\bibitem [{\citenamefont {Florkowski}\ and\ \citenamefont
  {Maksymiuk}(2014)}]{Florkowski:2014sda}%
  \BibitemOpen
  \bibfield  {author} {\bibinfo {author} {\bibfnamefont {W.}~\bibnamefont
  {Florkowski}}\ and\ \bibinfo {author} {\bibfnamefont {E.}~\bibnamefont
  {Maksymiuk}},\ }\href@noop {} {\  (\bibinfo {year} {2014})},\ \Eprint
  {http://arxiv.org/abs/1411.3666} {arXiv:1411.3666 [hep-ph]} \BibitemShut
  {NoStop}%
\bibitem [{\citenamefont {Denicol}\ \emph
  {et~al.}(2014{\natexlab{b}})\citenamefont {Denicol}, \citenamefont {Heinz},
  \citenamefont {Martinez}, \citenamefont {Noronha},\ and\ \citenamefont
  {Strickland}}]{Denicol:2014xca}%
  \BibitemOpen
  \bibfield  {author} {\bibinfo {author} {\bibfnamefont {G.~S.}\ \bibnamefont
  {Denicol}}, \bibinfo {author} {\bibfnamefont {U.~W.}\ \bibnamefont {Heinz}},
  \bibinfo {author} {\bibfnamefont {M.}~\bibnamefont {Martinez}}, \bibinfo
  {author} {\bibfnamefont {J.}~\bibnamefont {Noronha}}, \ and\ \bibinfo
  {author} {\bibfnamefont {M.}~\bibnamefont {Strickland}},\ }\href {\doibase
  10.1103/PhysRevLett.113.202301} {\bibfield  {journal} {\bibinfo  {journal}
  {Phys.Rev.Lett.}\ }\textbf {\bibinfo {volume} {113}},\ \bibinfo {pages}
  {202301} (\bibinfo {year} {2014}{\natexlab{b}})},\ \Eprint
  {http://arxiv.org/abs/1408.5646} {arXiv:1408.5646 [hep-ph]} \BibitemShut
  {NoStop}%
\bibitem [{\citenamefont {Denicol}\ \emph
  {et~al.}(2014{\natexlab{c}})\citenamefont {Denicol}, \citenamefont {Heinz},
  \citenamefont {Martinez}, \citenamefont {Noronha},\ and\ \citenamefont
  {Strickland}}]{Denicol:2014tha}%
  \BibitemOpen
  \bibfield  {author} {\bibinfo {author} {\bibfnamefont {G.~S.}\ \bibnamefont
  {Denicol}}, \bibinfo {author} {\bibfnamefont {U.~W.}\ \bibnamefont {Heinz}},
  \bibinfo {author} {\bibfnamefont {M.}~\bibnamefont {Martinez}}, \bibinfo
  {author} {\bibfnamefont {J.}~\bibnamefont {Noronha}}, \ and\ \bibinfo
  {author} {\bibfnamefont {M.}~\bibnamefont {Strickland}},\ }\href {\doibase
  10.1103/PhysRevD.90.125026} {\bibfield  {journal} {\bibinfo  {journal}
  {Phys.Rev.}\ }\textbf {\bibinfo {volume} {D90}},\ \bibinfo {pages} {125026}
  (\bibinfo {year} {2014}{\natexlab{c}})},\ \Eprint
  {http://arxiv.org/abs/1408.7048} {arXiv:1408.7048 [hep-ph]} \BibitemShut
  {NoStop}%
\bibitem [{\citenamefont {Bozek}\ and\ \citenamefont
  {Wyskiel}(2009)}]{Bozek:2009ty}%
  \BibitemOpen
  \bibfield  {author} {\bibinfo {author} {\bibfnamefont {P.}~\bibnamefont
  {Bozek}}\ and\ \bibinfo {author} {\bibfnamefont {I.}~\bibnamefont
  {Wyskiel}},\ }\href {\doibase 10.1103/PhysRevC.79.044916} {\bibfield
  {journal} {\bibinfo  {journal} {Phys.Rev.}\ }\textbf {\bibinfo {volume}
  {C79}},\ \bibinfo {pages} {044916} (\bibinfo {year} {2009})},\ \Eprint
  {http://arxiv.org/abs/0902.4121} {arXiv:0902.4121 [nucl-th]} \BibitemShut
  {NoStop}%
\bibitem [{\citenamefont {Martinez}\ and\ \citenamefont
  {Strickland}(2009)}]{Martinez:2008mc}%
  \BibitemOpen
  \bibfield  {author} {\bibinfo {author} {\bibfnamefont {M.}~\bibnamefont
  {Martinez}}\ and\ \bibinfo {author} {\bibfnamefont {M.}~\bibnamefont
  {Strickland}},\ }\href {\doibase 10.1140/epjc/s10052-008-0851-8} {\bibfield
  {journal} {\bibinfo  {journal} {Eur.Phys.J.}\ }\textbf {\bibinfo {volume}
  {C61}},\ \bibinfo {pages} {905} (\bibinfo {year} {2009})},\ \Eprint
  {http://arxiv.org/abs/0808.3969} {arXiv:0808.3969 [hep-ph]} \BibitemShut
  {NoStop}%
\bibitem [{\citenamefont {Shuryak}(2012)}]{Shuryak:2012nf}%
  \BibitemOpen
  \bibfield  {author} {\bibinfo {author} {\bibfnamefont {E.}~\bibnamefont
  {Shuryak}},\ }\href@noop {} {\  (\bibinfo {year} {2012})},\ \Eprint
  {http://arxiv.org/abs/1203.1012} {arXiv:1203.1012 [nucl-th]} \BibitemShut
  {NoStop}%
\bibitem [{\citenamefont {Dion}\ \emph {et~al.}(2011)\citenamefont {Dion},
  \citenamefont {Paquet}, \citenamefont {Schenke}, \citenamefont {Young},
  \citenamefont {Jeon} \emph {et~al.}}]{Dion:2011pp}%
  \BibitemOpen
  \bibfield  {author} {\bibinfo {author} {\bibfnamefont {M.}~\bibnamefont
  {Dion}}, \bibinfo {author} {\bibfnamefont {J.-F.}\ \bibnamefont {Paquet}},
  \bibinfo {author} {\bibfnamefont {B.}~\bibnamefont {Schenke}}, \bibinfo
  {author} {\bibfnamefont {C.}~\bibnamefont {Young}}, \bibinfo {author}
  {\bibfnamefont {S.}~\bibnamefont {Jeon}},  \emph {et~al.},\ }\href {\doibase
  10.1103/PhysRevC.84.064901} {\bibfield  {journal} {\bibinfo  {journal}
  {Phys.Rev.}\ }\textbf {\bibinfo {volume} {C84}},\ \bibinfo {pages} {064901}
  (\bibinfo {year} {2011})},\ \Eprint {http://arxiv.org/abs/1109.4405}
  {arXiv:1109.4405 [hep-ph]} \BibitemShut {NoStop}%
\bibitem [{\citenamefont {Chatterjee}\ \emph {et~al.}(2011)\citenamefont
  {Chatterjee}, \citenamefont {Holopainen}, \citenamefont {Renk},\ and\
  \citenamefont {Eskola}}]{Chatterjee:2011dw}%
  \BibitemOpen
  \bibfield  {author} {\bibinfo {author} {\bibfnamefont {R.}~\bibnamefont
  {Chatterjee}}, \bibinfo {author} {\bibfnamefont {H.}~\bibnamefont
  {Holopainen}}, \bibinfo {author} {\bibfnamefont {T.}~\bibnamefont {Renk}}, \
  and\ \bibinfo {author} {\bibfnamefont {K.~J.}\ \bibnamefont {Eskola}},\
  }\href {\doibase 10.1103/PhysRevC.83.054908} {\bibfield  {journal} {\bibinfo
  {journal} {Phys.Rev.}\ }\textbf {\bibinfo {volume} {C83}},\ \bibinfo {pages}
  {054908} (\bibinfo {year} {2011})},\ \Eprint {http://arxiv.org/abs/1102.4706}
  {arXiv:1102.4706 [hep-ph]} \BibitemShut {NoStop}%
\bibitem [{\citenamefont {Shen}\ \emph {et~al.}(2013)\citenamefont {Shen},
  \citenamefont {Heinz}, \citenamefont {Paquet}, \citenamefont {Kozlov},\ and\
  \citenamefont {Gale}}]{Shen:2013cca}%
  \BibitemOpen
  \bibfield  {author} {\bibinfo {author} {\bibfnamefont {C.}~\bibnamefont
  {Shen}}, \bibinfo {author} {\bibfnamefont {U.~W.}\ \bibnamefont {Heinz}},
  \bibinfo {author} {\bibfnamefont {J.-F.}\ \bibnamefont {Paquet}}, \bibinfo
  {author} {\bibfnamefont {I.}~\bibnamefont {Kozlov}}, \ and\ \bibinfo {author}
  {\bibfnamefont {C.}~\bibnamefont {Gale}},\ }\href@noop {} {\  (\bibinfo
  {year} {2013})},\ \Eprint {http://arxiv.org/abs/1308.2111} {arXiv:1308.2111
  [nucl-th]} \BibitemShut {NoStop}%
\bibitem [{\citenamefont {Schenke}\ and\ \citenamefont
  {Strickland}(2007)}]{Schenke:2006yp}%
  \BibitemOpen
  \bibfield  {author} {\bibinfo {author} {\bibfnamefont {B.}~\bibnamefont
  {Schenke}}\ and\ \bibinfo {author} {\bibfnamefont {M.}~\bibnamefont
  {Strickland}},\ }\href {\doibase 10.1103/PhysRevD.76.025023} {\bibfield
  {journal} {\bibinfo  {journal} {Phys. Rev.}\ }\textbf {\bibinfo {volume}
  {D76}},\ \bibinfo {pages} {025023} (\bibinfo {year} {2007})},\ \Eprint
  {http://arxiv.org/abs/hep-ph/0611332} {arXiv:hep-ph/0611332} \BibitemShut
  {NoStop}%
\bibitem [{\citenamefont {Schenke}(2008)}]{Schenke:2008hw}%
  \BibitemOpen
  \bibfield  {author} {\bibinfo {author} {\bibfnamefont {B.}~\bibnamefont
  {Schenke}},\ }\emph {\bibinfo {title} {{Collective Phenomena in the
  Non-Equilibrium Quark-Gluon Plasma}}},\ \href@noop {} {Ph.D. thesis},\
  \bibinfo  {school} {Johann Wolfgang Goethe-University, Frankfurt am Main,
  Germany} (\bibinfo {year} {2008}),\ \Eprint {http://arxiv.org/abs/0810.4306}
  {arXiv:0810.4306 [hep-ph]} \BibitemShut {NoStop}%
\bibitem [{\citenamefont {Bhattacharya}\ and\ \citenamefont
  {Roy}(2008)}]{Bhattacharya:2008up}%
  \BibitemOpen
  \bibfield  {author} {\bibinfo {author} {\bibfnamefont {L.}~\bibnamefont
  {Bhattacharya}}\ and\ \bibinfo {author} {\bibfnamefont {P.}~\bibnamefont
  {Roy}},\ }\href {\doibase 10.1103/PhysRevC.78.064904} {\bibfield  {journal}
  {\bibinfo  {journal} {Phys.Rev.}\ }\textbf {\bibinfo {volume} {C78}},\
  \bibinfo {pages} {064904} (\bibinfo {year} {2008})},\ \Eprint
  {http://arxiv.org/abs/0809.4596} {arXiv:0809.4596 [hep-ph]} \BibitemShut
  {NoStop}%
\bibitem [{\citenamefont {Bhattacharya}\ and\ \citenamefont
  {Roy}(2009)}]{Bhattacharya:2008mv}%
  \BibitemOpen
  \bibfield  {author} {\bibinfo {author} {\bibfnamefont {L.}~\bibnamefont
  {Bhattacharya}}\ and\ \bibinfo {author} {\bibfnamefont {P.}~\bibnamefont
  {Roy}},\ }\href {\doibase 10.1103/PhysRevC.79.054910} {\bibfield  {journal}
  {\bibinfo  {journal} {Phys.Rev.}\ }\textbf {\bibinfo {volume} {C79}},\
  \bibinfo {pages} {054910} (\bibinfo {year} {2009})},\ \Eprint
  {http://arxiv.org/abs/0812.1478} {arXiv:0812.1478 [hep-ph]} \BibitemShut
  {NoStop}%
\bibitem [{\citenamefont {Bhattacharya}\ and\ \citenamefont
  {Roy}(2010{\natexlab{a}})}]{Bhattacharya:2009sb}%
  \BibitemOpen
  \bibfield  {author} {\bibinfo {author} {\bibfnamefont {L.}~\bibnamefont
  {Bhattacharya}}\ and\ \bibinfo {author} {\bibfnamefont {P.}~\bibnamefont
  {Roy}},\ }\href {\doibase 10.1103/PhysRevC.81.054904} {\bibfield  {journal}
  {\bibinfo  {journal} {Phys.Rev.}\ }\textbf {\bibinfo {volume} {C81}},\
  \bibinfo {pages} {054904} (\bibinfo {year} {2010}{\natexlab{a}})},\ \Eprint
  {http://arxiv.org/abs/0907.3607} {arXiv:0907.3607 [hep-ph]} \BibitemShut
  {NoStop}%
\bibitem [{\citenamefont {Bhattacharya}\ and\ \citenamefont
  {Roy}(2010{\natexlab{b}})}]{Bhattacharya:2010sq}%
  \BibitemOpen
  \bibfield  {author} {\bibinfo {author} {\bibfnamefont {L.}~\bibnamefont
  {Bhattacharya}}\ and\ \bibinfo {author} {\bibfnamefont {P.}~\bibnamefont
  {Roy}},\ }\href {\doibase 10.1088/0954-3899/37/10/105010} {\bibfield
  {journal} {\bibinfo  {journal} {J.Phys.G}\ }\textbf {\bibinfo {volume}
  {G37}},\ \bibinfo {pages} {105010} (\bibinfo {year} {2010}{\natexlab{b}})},\
  \Eprint {http://arxiv.org/abs/1001.1054} {arXiv:1001.1054 [hep-ph]}
  \BibitemShut {NoStop}%
\bibitem [{\citenamefont {McLerran}\ and\ \citenamefont
  {Schenke}(2014)}]{McLerran:2014hza}%
  \BibitemOpen
  \bibfield  {author} {\bibinfo {author} {\bibfnamefont {L.}~\bibnamefont
  {McLerran}}\ and\ \bibinfo {author} {\bibfnamefont {B.}~\bibnamefont
  {Schenke}},\ }\href {\doibase 10.1016/j.nuclphysa.2014.06.004} {\bibfield
  {journal} {\bibinfo  {journal} {Nucl.Phys.}\ }\textbf {\bibinfo {volume}
  {A929}},\ \bibinfo {pages} {71} (\bibinfo {year} {2014})},\ \Eprint
  {http://arxiv.org/abs/1403.7462} {arXiv:1403.7462 [hep-ph]} \BibitemShut
  {NoStop}%
\bibitem [{\citenamefont {Luzum}\ and\ \citenamefont
  {Romatschke}(2009)}]{Luzum:2009sb}%
  \BibitemOpen
  \bibfield  {author} {\bibinfo {author} {\bibfnamefont {M.}~\bibnamefont
  {Luzum}}\ and\ \bibinfo {author} {\bibfnamefont {P.}~\bibnamefont
  {Romatschke}},\ }\href {\doibase 10.1103/PhysRevLett.103.262302} {\bibfield
  {journal} {\bibinfo  {journal} {Phys.Rev.Lett.}\ }\textbf {\bibinfo {volume}
  {103}},\ \bibinfo {pages} {262302} (\bibinfo {year} {2009})},\ \Eprint
  {http://arxiv.org/abs/0901.4588} {arXiv:0901.4588 [nucl-th]} \BibitemShut
  {NoStop}%
\bibitem [{\citenamefont {Niemi}\ and\ \citenamefont
  {Denicol}(2014)}]{Niemi:2014wta}%
  \BibitemOpen
  \bibfield  {author} {\bibinfo {author} {\bibfnamefont {H.}~\bibnamefont
  {Niemi}}\ and\ \bibinfo {author} {\bibfnamefont {G.}~\bibnamefont
  {Denicol}},\ }\href@noop {} {\  (\bibinfo {year} {2014})},\ \Eprint
  {http://arxiv.org/abs/1404.7327} {arXiv:1404.7327 [nucl-th]} \BibitemShut
  {NoStop}%
\end{thebibliography}%

\end{document}